\documentclass[9pt,twocolumn]{extarticle}
\pdfoutput=1
\usepackage{soul}
\usepackage{titlesec}
\usepackage{cite}
\usepackage{booktabs}

\usepackage[utf8]{inputenc}
\usepackage{newunicodechar}
\newunicodechar{≈}{\approx}
\usepackage{dsfont}
\usepackage{graphicx}
\usepackage{subcaption}
\usepackage[margin=0.9in]{geometry}
\usepackage[usenames,dvipsnames]{color}
\usepackage[colorlinks,linkcolor=Blue,urlcolor=Blue,citecolor=Blue]{hyperref}
\usepackage{amsmath,amssymb}
\usepackage[squaren]{SIunits}

\usepackage{mathpazo}
\usepackage{courier}
\normalfont
\usepackage[T1]{fontenc}
\usepackage{caption}
\usepackage{upgreek}

\newcommand{\ad}[1]{\textsuperscript{#1}\kern-2pt}

\makeatletter
\def\blx@maxline{77}
\makeatother

\usepackage{capt-of}

\def\mytitle{High-pressure phase stability and superconductivity \\ in La-Zr-H hydrides}

\title{\vspace{-1.0cm}\Huge\textbf{\textrm{\mytitle}}}

\author{Ijaz Shahid,$^{1,2,\star}$ Maxim A. Grebeniuk,$^{3,\star}$ Jinbin Zhao,$^{4,\star}$ Ergen Bao,$^{1}$ Tianye Yu,$^{1}$ Xiangyang Liu,$^{1, 2}$ \\
Yi-Chi Zhang,$^{1, 2}$ Artem R. Oganov,$^{3}$ Yan Sun,$^{1, 2}$ Peitao Liu,$^{1,2,\dagger}$ and  Xing-Qiu Chen$^{1,2,\dagger}$ }

\date{}
\begin{document}
	\twocolumn[{
		\maketitle
		\vspace{-5mm}
		\begin{center}
			\begin{minipage}{1\textwidth}
				\begin{center}
					\textit{
						\textsuperscript{1} Shenyang National Laboratory for Materials Science, Institute of Metal Research, Chinese Academy of Sciences, Shenyang 110016, China
						\\\textsuperscript{2} School of Materials Science and Engineering, University of Science and Technology of China, Shenyang 110016, China
						\\\textsuperscript{3} Skolkovo Institute of Science and Technology, Bolshoy Boulevard 30, bld. 1, 121205, Moscow, Russia
						\\\textsuperscript{4} Department of Computer Science, Xinzhou Normal University, Xinzhou 034000, China
						\vspace{5mm}
						\\{$\dagger$} Corresponding to: ptliu@imr.ac.cn, xingqiu.chen@imr.ac.cn
						\\{$\star$} These authors contribute equally.
						\vspace{5mm}
					}
				\end{center}
			\end{minipage}
		\end{center}

		\setlength\parindent{13pt}
		\begin{quotation}
			\noindent
			\section*{Abstract}
Hydrogen-rich ternary hydrides are promising candidates for high-$T_c$ superconductivity at megabar pressures, yet their chemical space is vast and largely unexplored. 
Combining evolutionary structure searches with first-principles calculations, we comprehensively investigate the La-Zr-H ternary system in the 150-300~GPa pressure range. 
Zero-point energy-corrected convex hull analysis identifies multiple stable superconducting phases, including $R3m$-Zr$_{2}$H$_{17}$ at 300~GPa and $P6/mmm$-LaZr$_{2}$H$_{24}$ at 200~GPa, both of which are thermodynamically and dynamically stable and exhibit strong electron-phonon coupling.
Solution of the Eliashberg equations predicts high superconducting transition temperatures of $T_c$ = 209~K for $R3m$-Zr$_{2}$H$_{17}$ at 300~GPa and $T_c$ = 202~K for $P6/mmm$-LaZr$_{2}$H$_{24}$ at 200~GPa.
In addition to these stable phases, we identify a high-symmetry metastable compound, $P\bar{6}m2$-LaZrH$_{18}$, which lies just ~0.027~eV/atom above the convex hull yet remains dynamically stable and exhibits a high predicted $T_c$ of 206~K at 300~GPa.
We find that, across all phases, the elevated $T_c$ correlates with the high-symmetry structure with dense hydrogen cages, favorable electron counts per hydrogen, and a large hydrogen-derived density of states at the Fermi level.
Finally, a random-forest machine learning model, trained on diverse hydrides superconductivity data, 
reproduces these structure-property trends across predicted structures, 
enabling to identify potential hydrides with high predicted $T_c$ for targeted follow-up calculations and future high-pressure experiments.
	\end{quotation}
	}]

	\newpage
	\clearpage

\section*{INTRODUCTION}
Superconductivity continues to be a central research focus in condensed matter physics due to its unique zero-resistance and diamagnetic properties~\cite{de1996discovery, lilia20222021, mangin2016superconductivity}. These characteristics enable transformative applications in power transmission, medical imaging, and quantum sensing~\cite{hirsch2015superconducting}, thereby driving the search for superconductors with high $T_c$.
Unconventional copper-based superconductors have achieved $T_c$ ranging from 20 to 133 K at ambient pressure~\cite{bednorz1986possible, schilling1993superconductivity, keimer2015quantum, wu1987superconductivity}, while iron-based superconductors exhibit $T_c$ values between 26 and 77 K~\cite{wang2012interface, kamihara2008iron, paglione2010high}. Recently, the field has witnessed exciting developments in nickel-based superconductors~\cite{li2019superconductivity, sun2023signatures, zhu2024superconductivity, li2024signature}. 
However, the pairing mechanisms for unconventional superconductors remain actively debated~\cite{pickett2023colloquium}, which limits their predictive capabilities. In contrast, conventional superconductors, such as $\mathrm{Mg}\mathrm{B}_2$~\cite{nagamatsu2001superconductivity}, benefit from the well-understood Bardeen-Cooper-Schrieffer (BCS) theory~\cite{bardeen1957microscopic}, allowing for predictable values of $T_c$. This predictability has spurred a data-driven revolution in discovery, encompassing high-throughput screening of known materials~\cite{sb28-fjc9}, BCS-inspired multi-step workflows accelerated by deep learning~\cite{choudhary2022}, large-scale {\it ab initio} landscape surveys~\cite{PhysRevB.104.054501}, and focused mapping of hydride phase spaces via template-based databases~\cite{PhysRevMaterials.7.054806}. Notably, a recent deep-learning framework scanned ~36 million ternary hydrides, predicting 144 high-$T_c$ candidates ($>$200 K) and uncovering 27 new structural prototypes~\cite{wang2026computational}.

The success of the BCS theory has propelled the emergence of polyhydrides-hydrogen-rich compounds containing exceptionally high hydrogen concentrations---as the current record-holders for high-temperature superconductivity. Polyhydrides leverage quantum effects in dense hydrogen lattices to achieve superconductivity at previously unattainable temperatures, offering new pathways to overcome the fundamental limitations of existing superconducting materials. 
Ashcroft's hypothesis laid the foundation for this discovery, proposing that hydrogen-rich materials could exhibit superconductivity under high-pressure conditions~\cite{ashcroft1968metallic}. This idea arises from the unique properties of hydrogen, the lightest element, which, in its molecular form, has been predicted to transition into a metallic phase when subjected to extreme pressures. Such a transition, according to the BCS theory, could result in atomic metallic hydrogen exhibiting strong electron-phonon coupling (EPC) and a high Debye temperature, both of which are key factors for potential high-temperature superconductivity~\cite{bardeen1957microscopic, dias2017observation}. 

Based on the Ashcroft's hypothesis~\cite{ashcroft1968metallic}, hydrogen-rich metallic systems can achieve high $T_c$ at comparatively lower pressures. In this regard, various hydrides have been identified to exhibit remarkable superconductivity, including but not limited to $Im\bar{3}m$-H$_{3}$S ($T_c\sim$203 K)~\cite{drozdov2015conventional}, $Cmca$-$\mathrm{H_2S}$ ($T_c\sim$ 80 K)~\cite{li2014metallization},  {{$Im\bar{3}m$-$\left(\mathrm{H_{2}S}\right)_{2}\mathrm{H_{2}}$ ($T_c\sim$ 191-204 K)~\cite{duan2014pressure},
$Im\bar{3}m$-$\mathrm{YH_{6}}$ ($T_c\sim$ 224 K)~\cite{troyan2021anomalous}, 
$Im\bar{3}m$-$\mathrm{CaH_{6}}$ ($T_c\sim$ 215 K)~\cite{ma2022high}, $Fm\bar{3}m$-$\mathrm{LaH_{10}}$ ($T_c\sim$ 250 K)~\cite{somayazulu2019evidence, drozdov2019superconductivity}, 
$R\bar{3}m$-$\mathrm{AcH_{10}}$ ($T_c\sim$ 204-251 K)~\cite{semenok2018actinium}, and $Fm\bar{3}m$-$\mathrm{ThH_{10}}$
($T_c\sim$ 159-161 K)~\cite{kvashnin2018high}. 
While these achievements require extreme pressures, a primary goal is stabilization at ambient conditions. Toward this goal, recent machine-learning-assisted screening of over 150,000 compounds has identified $\approx 50$ hydride systems with $T_c$ > 20 K (some > 70 K) that are candidates for ambient-pressure superconductivity, albeit often with slight thermodynamic instability~\cite{cerqueira2024searching}.

Binary hydrides exhibiting high $T_c$ typically fall within the "lability belt" of the Mendeleev's table~\cite{semenok2020distribution}.
Ternary hydrides, created by adding new elements or alloying to binary hydrides, often exhibit higher $T_c$ and more favorable stability at lower pressures~\cite{hilleke2022,liang2024,GENG2024101443}. For instance: ternary hydrides $\mathrm{H_9S_2In}$, $\mathrm{H_9S_2Sn}$, $\mathrm{H_9S_2Te}$, $\mathrm{H_6SIn}$ and $\mathrm{NaY_3H_{20}}$ were found to be dynamically stable at 20 GPa, with $T_c$ values of 54, 72, 5, 54 and 115 K, respectively~\cite{yang2026high,an2025design}. Similarly, $(\mathrm{La}, \mathrm{Y})\mathrm{H_{10}}$ ($T_c\sim$ 253 K)~\cite{semenok2021superconductivity}, Na$_{10}$Li$_{5}$H$_{86}$ ($T_c\sim$ 341 K) and CaLi$_{2}$H$_{17}$ ($T_c\sim$ 326 K)~\cite{ma2025high}, $\mathrm{Li_2NaH_{17}}$ ($T_c\sim$ 297 K), $\mathrm{ThY_2H_{24}}$ ($T_c\sim$ 303 K)~\cite{PhysRevB.111.054505}, $\mathrm{LaSc_2H_{24}}$ ($T_c\sim$ 271-298 K)~\cite{song2025}, $\mathrm{Th_2BH_{16}}$ ($T_c\sim$ 102 K)~\cite{gnq4-b1gr}, $\mathrm{S_7PH_{24}}$ ($T_c\sim$ 183 K)~\cite{doi:10.1021/acs.jpcc.1c10976} and $\mathrm{Li_2CaH_{17}}$ ($T_c\sim$ 370 K) have been reported~\cite{doi:10.1021/acs.chemmater.4c01472}, {(Lu,Y)$_4$H$_{23}$ ($T_c\sim$ 112 K at 215 GPa)~\cite{Zhang2025_A15_LuY4H23}, YBeH$_8$ ($T_c\sim$ 201 K at 200 GPa)~\cite{Du2024_YBeH8_JCP}, Sc$_2$MgH$_{18}$ ($T_c\sim$ 112 K at 150 GPa)~\cite{qin2025first}, YMgH$_{12}$ ($T_c = 190$~K at 200 GPa)~\cite{song2022systematic}. Among them, lanthanum-based ternary hydrides such as La-B-H~\cite{dicataldo2021labh8}, La-Ce-H~\cite{bi2022giant}, and $\mathrm{LaBeH_{8}}$~\cite{zhang2022design} also demonstrate the ability to stabilize at lower pressures, with $\mathrm{LaBeH_{8}}$ maintaining dynamic stability at just 20 GPa while exhibiting a remarkably high $T_c$ of 185 K~\cite{zhang2022design}. As illustrated above, lanthanum and yttrium-based hydrides have demonstrated impressive critical temperatures~\cite{somayazulu2019evidence, drozdov2019superconductivity, kong2019superconductivity}. Given its position adjacent to yttrium in the periodic table and its location within the "lability belt" of Mendeleev's table, zirconium containing hydrides seem potential candidates. It has exhibited encouraging $T_c$ values in binary hydrides~\cite{li2017phase, xie2020superconducting, abe2018high} and shares the relevant $d^{1}$, $d^{0}$, and $s^{2}d^{2}$ electronic configurations typical of high-$T_c$ phases~\cite{semenok2020distribution}.
These factors collectively motivate our investigation into the ternary La-Zr-H system for its potential to exhibit high-temperature superconductivity.

In this study, we systematically explored the chemical space of the La-Zr-H ternary system across a range of high pressures using an evolutionary structure search algorithm USPEX~\cite{lyakhov2013new, oganov2006crystal} combined with first-principles calculations.
We established the phase diagram of the La-Zr-H system and identified two thermodynamically stable superconducting phases, namely $R3m$-Zr$_{}$H$_{17}$ at 300 GPa and the high-symmetry $P6/mmm$-LaZr$_{2}$H$_{24}$ at 200 GPa, exhibiting strong electron-phonon coupling and high superconducting transition temperatures of 209 K and 202 K, respectively. 
In addition, we predicted a low-enthalpy metastable phase $P\bar{6}m2$-LaZrH$_{18}$ at 300 GPa, which remains dynamically stable and shows a high $T_c$ of 206 K.

\section*{RESULTS}
\noindent\textbf{Structure search and phase stability}\\
Evolutionary structure searches for the binary La-H and Zr-H systems at 0 and 200 GPa identified several stable and low-enthalpy metastable hydrides (see Supplementary Table 1). These binary results served as seeds for subsequent extensive searches in the ternary La-Zr-H system at 150, 200, 250, and 300 GPa (see ``Methods").  Previous study showed that the zero-point energy (ZPE) can affect the positioning of phases on the convex hull; for example, $\mathrm{ThH_9}$ becomes thermodynamically stable only after including the ZPE effects~\cite{Semenok2020}. To ensure reliable phase stability predictions in our ternary system, ZPE corrections were applied across all pressure ranges considered (see Fig.~1). For comparison, the non-ZPE-corrected convex hulls were also constructed (see Supplementary Fig.~1). The ternary structure searches were performed at pressures ranging from 150 to 300 GPa. The phase stability, defined as a compound's resistance to decomposition, was assessed by constructing the convex hull of formation enthalpy as a function of composition~\cite{Pickard2011}. Structures lying on the convex hull are thermodynamically stable, whereas those slightly above it are considered metastable provided they are dynamically stable, as indicated by the absence of imaginary phonon frequencies (see Supplementary Figs.~2–5).

\begin{figure*}[ht!]
\centering
\includegraphics[width=0.8\textwidth]{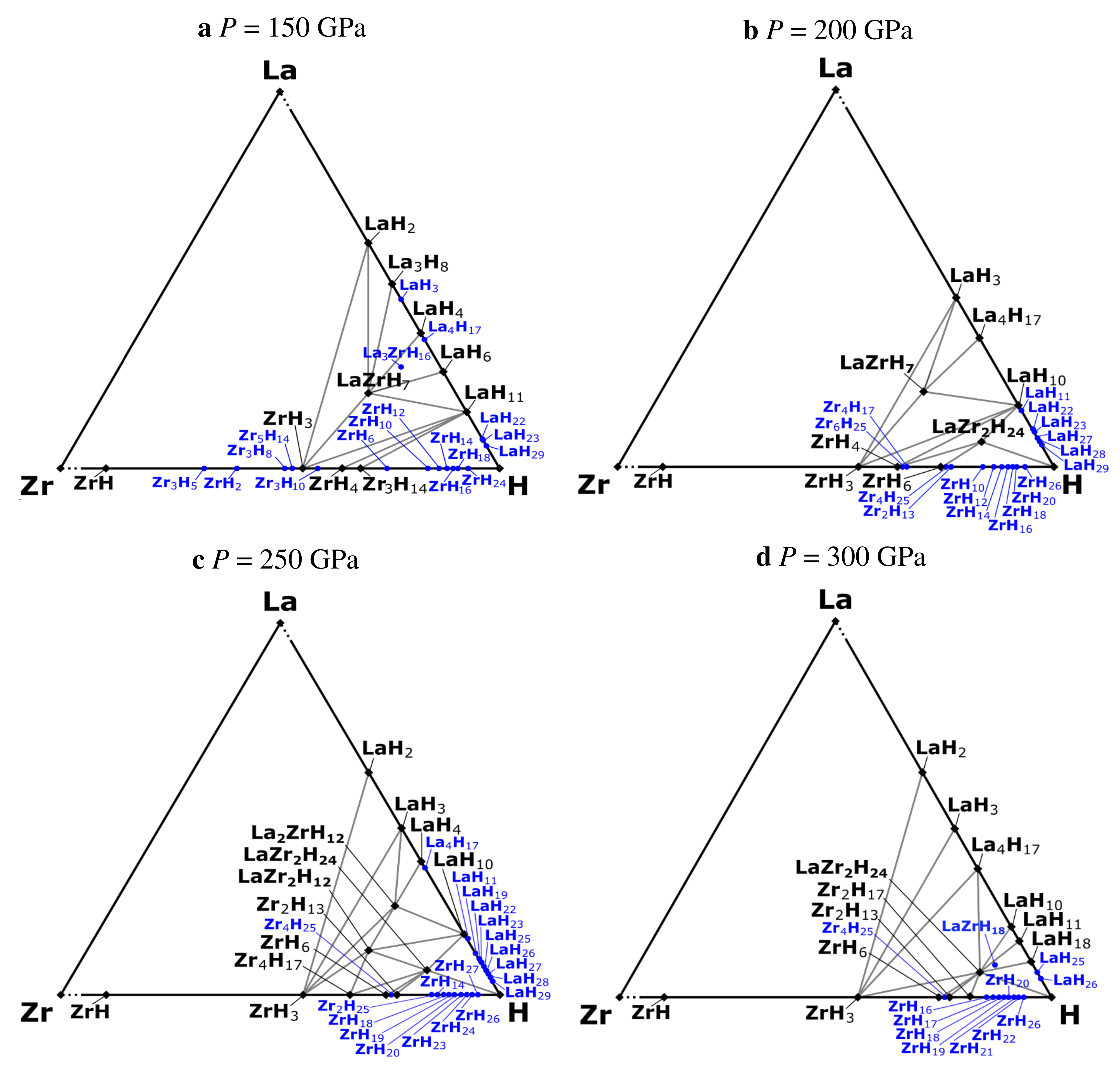}
\caption{\textbf{The ZPE-corrected convex hulls and phase diagrams for the La-Zr-H system at} \textbf {a}. 150 GPa \textbf{b}. 200 GPa \textbf{c}. 250 GPa and \textbf{d}. 300 GPa. Black diamonds represent thermodynamically stable phases, while blue circles denote metastable structures with formation enthalpies within 30 meV/atom above the convex hull.
		}
\label{Figure1}
\end{figure*}
Our structure searches revealed several novel ternary stable and metastable phases at multiple pressures, including $P4/mmm$-LaZrH$_{7}$ and a metastable $Pmm2$-La$_{3}$ZrH$_{16}$ at 150 GPa, $I4/mmm$-La$_{2}$ZrH$_{12}$ and $I4/mmm$-LaZr$_2$H$_{12}$ at 250 GPa, the thermodynamically stable $P6/mmm$-LaZr$_2$H$_{24}$ at 200 GPa, and a low-enthalpy metastable phase $P\bar{6}m2$-LaZrH$_{18}$ at 300 GPa (see Fig.~1).

At 200 GPa, three ternary compounds were identified: $I4/mmm$-La$_{2}$ZrH$_{12}$, $P4/mmm$-LaZrH$_{8}$, and $Pmm2$-La$_{3}$ZrH$_{16}$. However, phonon calculations indicate that all three phases are dynamically unstable (see Supplementary Fig.~6) and were therefore not considered in the subsequent analysis. In addition to these ternary compounds, several previously unreported binary hydrides were also identified, e.g., $C2/m$-La$_{4}$H$_{17}$, $Cmmm$-Zr$_{3}$H$_{14}$, and $Immm$-La$_{3}$H$_{8}$ at 150 GPa, $Cm$-Zr$_{2}$H$_{13}$, $I4/mmm$-LaZr$_{2}$H$_{12}$ and $I4/mmm$-La$_{2}$ZrH$_{12}$ at 250 GPa, $R3m$-Zr$_{2}$H$_{17}$ and $P\bar6m2$-LaZrH$_{18}$ at 300 GPa. The structural parameters of these phases are summarized in Supplementary Table 2.\\

\noindent\textbf{Screening for potential high-$T_c$ structures}\\
\noindent Based on previous studies~\cite{xie2020hydrogen,shutov2024ternary}, high-symmetry, hydrogen-rich crystal structures (referred to as criterion A), together with an favorable electron transfer per hydrogen atom ($\bar e/\mathrm{H}$$\sim$0.33)  (called criterion B), are favorable for achieving high superconducting transition temperatures. When $\bar e/\mathrm{H}$ is small (approximately 0-0.13), hydrogen atoms tend to form H-H molecular units, resulting in molecular hydrides that generally exhibit low or negligible $T_c$. In contrast, large $\bar e/\mathrm{H}$ values (approaching $\sim$1) correspond to strongly ionic hydrides dominated by $\mathrm{H}^{\rm-}$ species, and may lead to non-superconducting properties. Notably, the highest $T_c$ values are observed at intermediate $\bar e/\mathrm{H}$ values around 0.33, where hydrogen forms extended, metallic networks that are conducive to strong electron-phonon coupling. This trend has been consistently reported in several high-$T_c$ binary and ternary hydrides, including LaH$_{10}$~\cite{somayazulu2019evidence,drozdov2019superconductivity},
YH$_9$~\cite{kong2019superconductivity}, AcH$_{10}$~\cite{semenok2018actinium},
LaMg$_3$H$_{28}$~\cite{shutov2024ternary}, and YZrH$_{18}$~\cite{Zhao2023_YZrH},
all of which exhibit $\bar e/\mathrm{H}$ values in the range of approximately 0.30--0.38.

\begin{figure*}[ht!]
\centering
\includegraphics[width=0.8\textwidth]{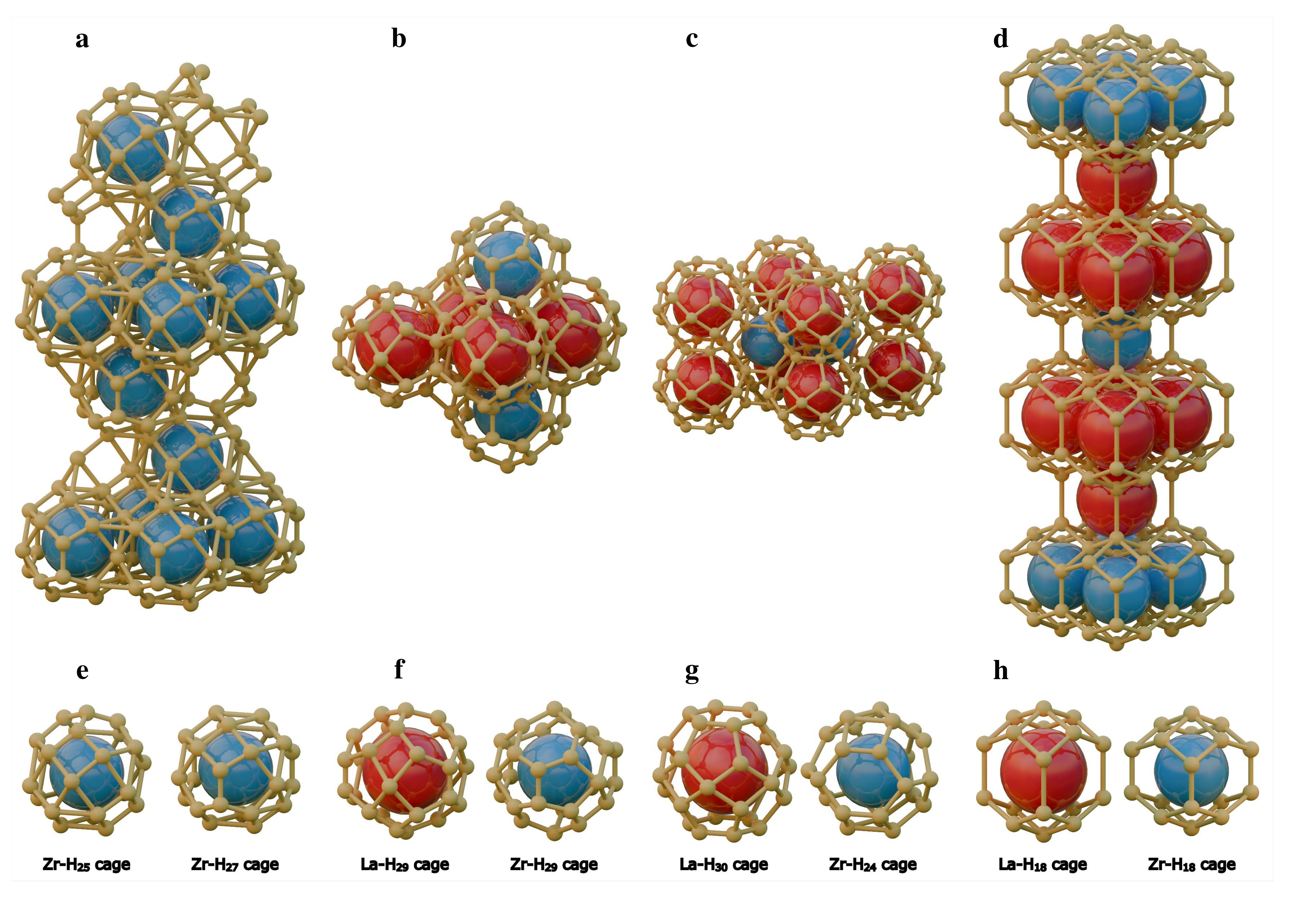}
\caption{\textbf{Crystal structures.} \textbf {a}. $R3m$-Zr$_{2}$H$_{17}$.
 \textbf{b}. $P\bar6m2$-LaZrH$_{18}$. \textbf{c}. $P6/mmm$-LaZr$_{2}$H$_{24}$. 
 \textbf{d}. $I4/mmm$-La$_{2}$ZrH$_{12}$. \textbf {e}. Zr-H$_{25}$ and Zr-H$_{27}$ cage in $R3m$-Zr$_{2}$H$_{17}$. \textbf {f}. La-H$_{29}$ and  Zr-H$_{29}$ cages in $P\bar6m2$-LaZrH$_{18}$. \textbf{g}. La-H$_{30}$ and Zr-H$_{24}$ cages in $P6/mmm$-LaZr$_{2}$H$_{24}$. \textbf{h}. La-H$_{18}$ and Zr-H$_{18}$ cages in $I4/mmm$-La$_{2}$ZrH$_{12}$.
 Red, blue, and golden yellow spheres represent La, Zr, and H atoms, respectively.
		}
\label{Figure2}
\end{figure*}

\begin{figure*}[ht!]
\centering
\includegraphics[width=0.8\textwidth]{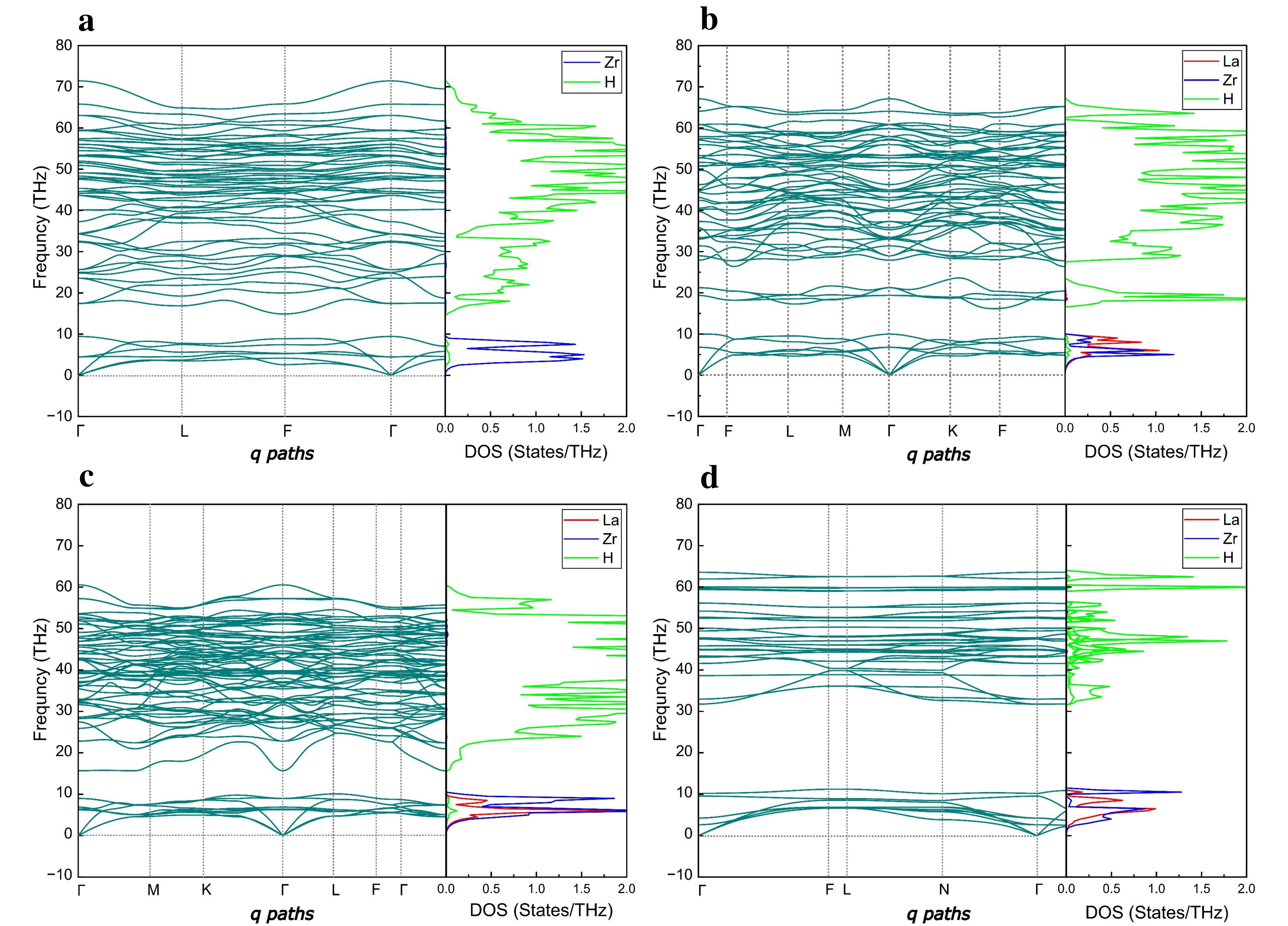} 
\caption{\textbf{Phonon dispersion relationships and partial phonon DOSs.} 
\textbf {a}. $R3m$-Zr$_{2}$H$_{17}$ at 300 GPa.
\textbf{b}. $P\bar6m2$-LaZrH$_{18}$ at 300 GPa.
\textbf{c}. $P6/mmm$-LaZr$_{2}$H$_{24}$ at 200 GPa.
\textbf{d}. $I4/mmm$-La$_{2}$ZrH$_{12}$ at 250 GPa.
		}
\label{Figure3}
\end{figure*}

Based on criteria A and B, we selected $R3m$-Zr$_2$H$_{17}$ ($\bar e$/H = 0.47), $P\bar6m2$-LaZrH$_{18}$ ($\bar e$/H = 0.38) at 300 GPa and $P6/mmm$-LaZr$_{2}$H$_{24}$ ($\bar e$/H = 0.46) at 200 GPa as representative candidates that satisfy both criteria. To provide a meaningful contrast and further evaluate the robustness of criterion B, the high-symmetry phase $I4/mmm$-La$_{2}$ZrH$_{12}$ was also included, despite its elevated $\bar{e}$/H value of 0.83. This selection enables a direct comparison between compounds that closely follow or deviate from the proposed criteria, thereby facilitating a clearer evaluation of the structure–property relationships explored in this study.\\

\begin{figure*}[ht!]
\centering
\includegraphics[width=0.8\textwidth]{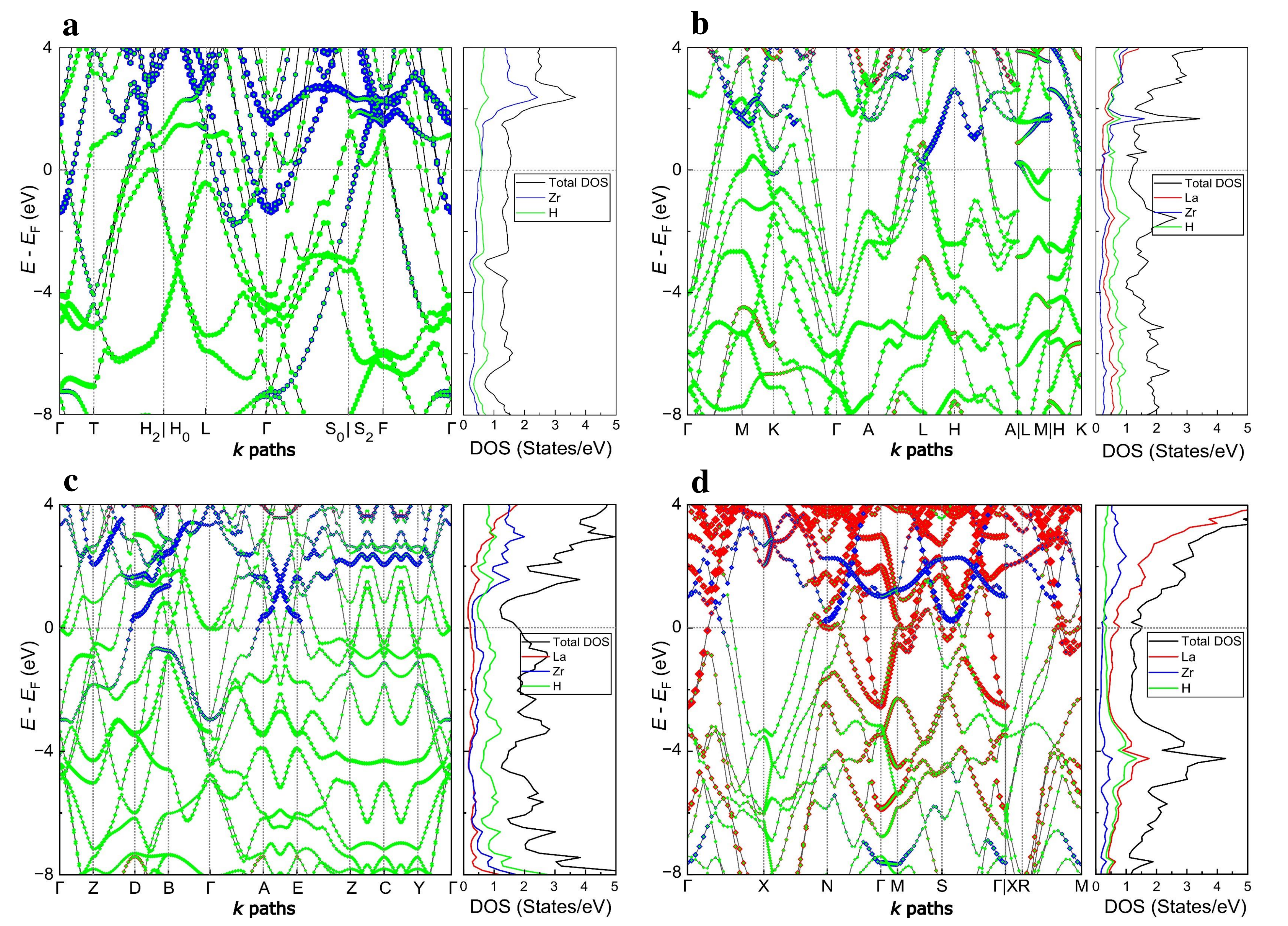}
\caption{\textbf{Electronic band structures and DOSs.} 
\textbf {a}. $R3m$-Zr$_{2}$H$_{17}$ at 300 GPa.
\textbf{b}. $P\bar6m2$-LaZrH$_{18}$ at 300 GPa.
\textbf{c}. $P6/mmm$-LaZr$_{2}$H$_{24}$ at 200 GPa.
\textbf{d}. $I4/mmm$-La$_{2}$ZrH$_{12}$ at 250 GPa.
		}
\label{Figure4}
\end{figure*}

\noindent\textbf{Structural properties of selected compounds}\\
\noindent The crystal structures of four selected representative compounds are shown in Fig.~\ref{Figure2}. The hydrogen-rich $R3m$-Zr$_2$H$_{17}$, $P\bar6m2$-LaZrH$_{18}$ phases (both at 300 GPa) and $P6/mmm$-LaZr$_{2}$H$_{24}$ at 200 GPa exhibit relatively high crystallographic symmetry, whereas $I4/mmm$-La$_{2}$ZrH$_{12}$ (250 GPa) is distinguished by its reduced hydrogen content, enabling a systematic comparison of structural motifs and hydrogen sublattice connectivity. $R3m$-Zr$_2$H$_{17}$ evolves from being metastable at 150 GPa (54 meV/atom above the convex hull) to thermodynamically stable at 300 GPa (Fig.~\ref{Figure1}d). In this phase, Zr atoms are embedded in a densely connected hydrogen framework with an average H-H separation of 1.28~\AA, exceeding the molecular H$_2$ bond length and indicating an extended hydrogen network. In hydrogen-rich compounds, the unique hydrogen cages are known to enhance electron-phonon interactions by modifying the electronic states near the Fermi level~\cite{Wang2025}. In $R3m$-Zr$_{2}$H$_{17}$, the Zr atoms are surrounded by two types of densely packed H-cages with high coordination number of 25 and 27 (Fig.~\ref{Figure2}e), which not only contributes to structural stability under high pressure but also creates a favorable environment for superconductivity. This stability is further reinforced by a large electronegativity difference between Zr (1.33) and H (2.18), which drives electron transfer from Zr to H, intensifying under compression as interatomic distances shrink. Notably, studies on sulfur hydride~\cite{drozdov2015conventional} also demonstrated that such hydrogen-rich frameworks facilitate strong EPC and high-frequency phonon modes, which are key factors for high-$T_c$ superconductivity. 
The ternary phase $P\bar6m2$-LaZrH$_{18}$ lies slightly above the convex hull (0.027 eV/atom at 300 GPa) yet remains dynamically stable and consists of La and Zr atoms enclosed within H$_{29}$ cages, forming a stacking of $\mathrm{La\text{-}H_{29}}$ and $\mathrm{Zr\text{-}H_{29}}$ cages (Fig.~\ref{Figure2}f) the average H-H distance is 1.13~\AA. This structure preserves a hydrogen-rich, highly coordinated cage motif (La/Zr: 29) while introducing chemical complexity through the mixed La-Zr framework.
Similarly, $P6/mmm$-LaZr$_{2}$H$_{24}$ phase at 200 GPa also exhibits a hydrogen-rich cage network composed of $\mathrm{La\text{-}H_{30}}$ and $\mathrm{Zr\text{-}H_{24}}$ (Fig.~\ref{Figure2}g). The average H–H separation is (1.26~\AA), which is similar to the above-discussed two phases at 300 GPa. The structure maintains dense hydrogen coordination around both metal species, preserving the key structural characteristics associated with strong electron-phonon coupling in high-pressure superhydrides. 
In contrast, the $I4/mmm$-La$_2$ZrH$_{12}$ phase, despite its high symmetry, exhibits a reduced coordination number of 18 (Fig.~\ref{Figure2}h) and lower hydrogen content. The corresponding average H-H distances increase to 1.47~\AA. Shorter H-H distances, often observed in hydrogen-rich phases, have been discussed to contribute to metallic hydrogen-like bonding networks~\cite{Sun2021LaSuperhydride}, which are favorable for enhanced superconducting properties. In this context, the relatively lower hydrogen content and longer H-H separations in $I4/mmm$-La$_{2}$ZrH$_{12}$ may account for its reduced superconducting transition temperatures, as will be demonstrated later.\\

\begin{figure*}[ht!]
\centering
\includegraphics[width=0.9\textwidth]{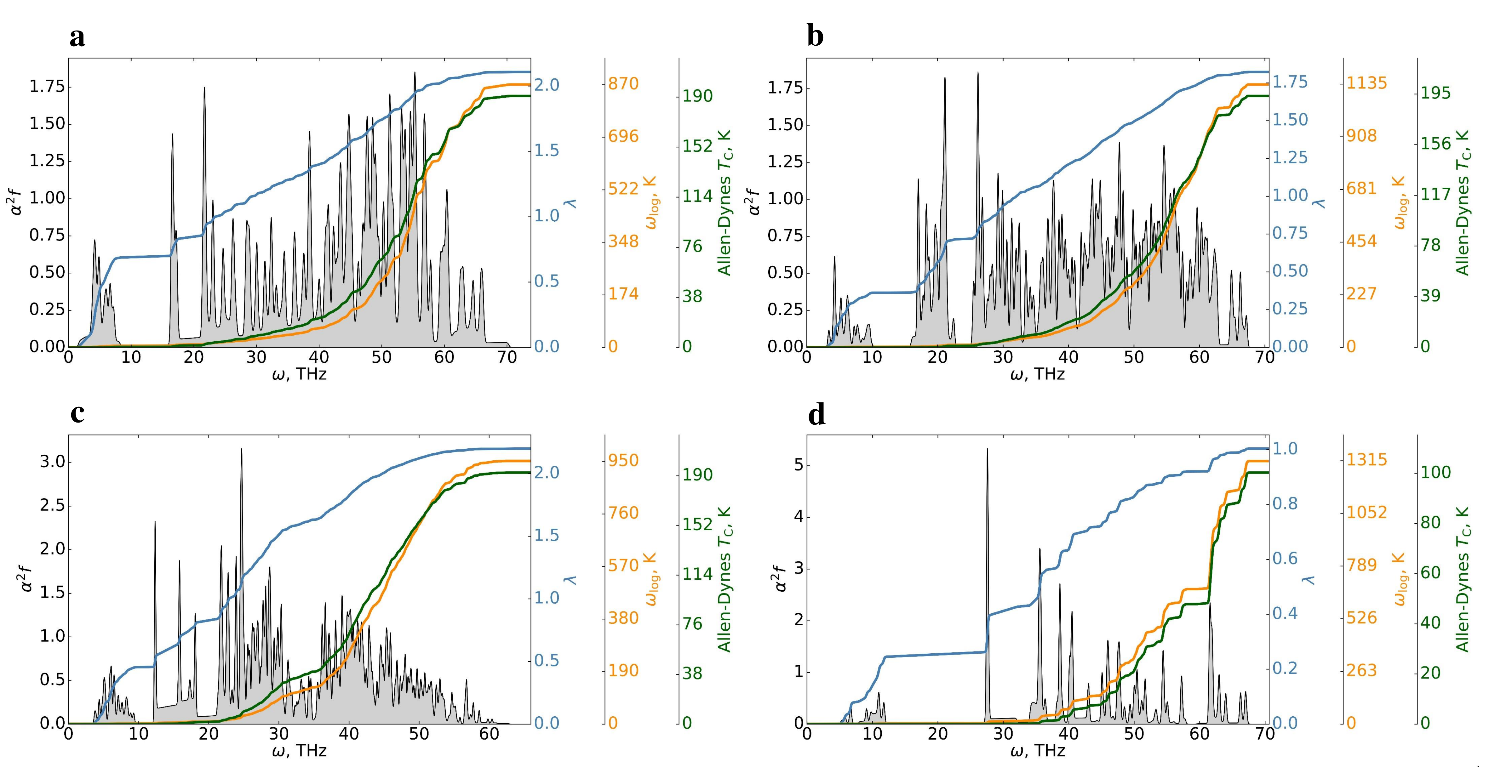}
\caption{\textbf{Eliashberg spectral function ($\alpha^2F$), EPC parameter ($\lambda$), logarithmic average phonon frequency ($\omega_{\log}$), and estimated critical transition temperature ($T_c$).} 
\textbf {a}. $R3m$-Zr$_{2}$H$_{17}$ at 300 GPa.
\textbf{b}. $P\bar6m2$-LaZrH$_{18}$ at 300 GPa.
\textbf{c}. $P6/mmm$-LaZr$_{2}$H$_{24}$ at 200 GPa.
\textbf{d}. $I4/mmm$-La$_{2}$ZrH$_{12}$ at 250 GPa.
		}
\label{Figure5}
\end{figure*}

\setlength{\fboxrule}{1.1pt}   
\setlength{\fboxsep}{4pt}     
\setlength{\arrayrulewidth}{1.0pt}  

\begin{table*}[t]
\centering
\label{tab:superconductivity-summary}
\renewcommand{\arraystretch}{1.3}

\fbox{%
\begin{minipage}{\textwidth}
\small
\textbf{Table 1.}
$T_c$ calculated using the McMillan and Allen-Dynes formulas, and Eliashberg equations. For the McMillan and Allen-Dynes approach, standard Coulomb pseudopotential values of $\mu^{*}_{\mathit{McMillan}}$ = 0.10 and 0.15 were used, whereas for the numerical solution of isotropic Eliashberg equations, rescaled values $\mu^{*}_{\mathit{Eliashberg}}$ (ranging from 0.12 to 0.22 depending on the system) were employed. The maximum phonon frequency ($\omega_{\mathit{ph}}$) and highest Matsubara frequency $\omega_{\mathit{max}}^{\mathit{Matsub}}$ used in the rescaling procedure are presented. $T_c$ values predicted by machine learning (ML) or empirical models are also included in this table. Additional parameters include EPC ($\lambda$), logarithmic average phonon frequency ($\omega_{\log}$), second moment of the normalized weight function $(\bar{\omega}_2)$, zero-temperature superconducting gap ($\Delta(0)$), thermodynamic critical magnetic field ($H_c(0)$), density of states ($N_f$), at the Fermi energy, Sommerfeld coefficient ($\gamma$), the H fraction (HDOS fraction) at the Fermi energy, and electron transfer per H atom ($\bar{e}$/H) for the four hydride compounds under different pressures.\\[4pt]
\noindent\rule{\textwidth}{1.5pt}  
\resizebox{\textwidth}{!}{%
\begin{tabular}{l c c c c}
\toprule
\textbf{Parameters} &
\textit{R${3}$m}-Zr$_{2}$H$_{17}$ (300 GPa) &
\textit{P$\bar{6}m2$}-LaZrH$_{18}$ (300 GPa) &
\textit{$P6/mmm$}-LaZr$_{2}$H$_{24}$ (200 GPa) &
\textit{I4/mmm}-La$_{2}$ZrH$_{12}$ (250 GPa) \\
\hline
$T_c$ (McMillan), K & 129 - 115 & 151 - 133 & 143 - 129 & 90 - 69  \\
$T_c$ (Allen-Dynes), K & 188 - 158 & 193 - 162 & 191 - 163 & 98 - 74  \\
$\mu^{*}_{\mathit{McMillan}}$ & 0.10 - 0.15 & 0.10 - 0.15 & 0.10 - 0.15 & 0.10 - 0.15  \\
$T_c$ (Eliashberg), K  & 209 - 178 & 206 - 173 & 202 - 175 & 97 - 75  \\
$\mu^{*}_{\mathit{Eliashberg}}$ & 0.13 - 0.21 & 0.13 - 0.22 & 0.13 - 0.22 & 0.12 - 0.18  \\
$\omega_{\mathit{ph}}$, THz & 70.29 & 67.27 & 62.79 & 67.27  \\
$\omega_{\mathit{max}}^{\mathit{Matsub}}$, THz & 590.96 - 496.62 & 606.67 - 509.19 & 600.38 - 512.33 & 308.01 - 232.51 \\
$T_c$ (ML model), Zhao, K~\cite{Zhao2024} & 187 & 244 & 229 & 71 \\
$T_c$ (Empirical model), Belli, K~\cite{belli2021strong} & 241 & 228 & 170 & 50 \\
$\lambda$ & 2.11 & 1.84 & 2.20 & 1.00 \\
$\omega_{\log}$, K  & 841 & 1111 &  941 & 1283 \\
$\bar{\omega}_2$, K & 1613 & 1690 & 1348 & 1788 \\
$\Delta(0)$, meV & 49.10 - 40.30 & 44.70 - 35.70 & 45.60 - 38.10 & 16.80 - 12.50  \\
$H_c(0)$, T & 5.30 - 4.60 & 4.30 - 3.60 & 4.20 - 3.60 & 1.50 - 1.10  \\
$N_f$, $\mathrm{states}/(\mathrm{eV}\cdot\text{\AA}^3)$ & 0.034 & 0.025 & 0.022 & 0.022  \\
$\gamma,\ \mathrm{\frac{mJ}{cm^3\cdot K^2}}$ & 0.4182 & 0.2813 & 0.2746 & 0.1793  \\
HDOS fraction & 0.69 & 0.40 & 0.40 & 0.20  \\
$\bar{e}$/H & 0.47 & 0.38 & 0.46 & 0.83  \\
\bottomrule
\end{tabular}
}
\end{minipage}
}
\end{table*}

\noindent \textbf{Dynamical stability}\\
\noindent The calculated phonon dispersions confirm the dynamical stability under the corresponding pressures for all four phases, as evidenced by the absence of imaginary phonon branches in Fig.~\ref{Figure3}. The partial phonon density of states (DOSs) further indicates that the low-frequency region is dominated by the vibrations of the heavier La/Zr sublattices, whereas the high-frequency region is primarily contributed by lighter hydrogen atoms, reflecting the strong mass contrast in compressed hydrides. A notable difference is observed in the accessible frequency range: the hydrogen-rich cage phases ($R3m$-Zr$_2$H$_{17}$ and $P\bar6m2$-LaZrH$_{18}$) exhibit H-derived phonon modes extending to higher frequencies (reaching $\sim$ 65-70~THz in Fig.~\ref{Figure3}a,b). By contrast, $P6/mmm$-LaZr$_2$H$_{24}$ shows a lower upper-frequency cutoff of about $\sim 60$~THz (Fig.~\ref{Figure3}c), while $I4/mmm$-La$_2$ZrH$_{12}$ extends slightly higher to $\sim 65$~THz (Fig.~\ref{Figure3}d). This indicates that LaZr$_2$H$_{24}$ possesses a somewhat softened hydrogen vibrational spectrum relative to the phases at 300 GPa, yet it still maintains a broad H-dominated phonon manifold capable of supporting strong electron-phonon coupling. In particular, $I4/mmm$-La$_2$ZrH$_{12}$ displays a more pronounced separation between the low-frequency metal-dominated phonon modes and the high-frequency H-dominated phonon modes, which may redistribute the phonon contributions relevant to electron-phonon coupling across different frequency regions.\\

\noindent \textbf{Electronic structure}\\
\noindent Figure~\ref{Figure4} shows the electronic band structures and the element-resolved DOSs for the four selected systems. One can see that all four phases are metallic,  exhibiting numerous band crossings at the Fermi energy along the high-symmetry paths. Notably, $R3m$-Zr$_2$H$_{17}$, $P\bar6m2$-LaZrH$_{18}$ and $P6/mmm$-LaZr$_{2}$H$_{24}$ exhibit a more pronounced H contribution to the total DOSs at $E_F$ (Fig.~\ref{Figure4}a,b and c), 
reaching 0.69, 0.40 and 0.40 (see Table 1).
Such enhanced H participation at $E_F$ is often associated with stronger electron-phonon coupling in high-pressure hydrides~\cite{Zhao2024}.
By contrast, the H contribution to the total DOSs at $E_F$ is comparatively reduced for $I4/mmm$-La$_2$ZrH$_{12}$ (Fig.~\ref{Figure4}d, Table 1), whose near-$E_F$ DOSs are more strongly dominated by metal-derived states. The charge analysis (Table~1) reveal values of $\bar e/\mathrm{H}$ to be 0.47 for $R3m$-Zr$_2$H$_{17}$, 0.38 for $P\bar6m2$-LaZrH$_{18}$ and 0.46 for $P6/mmm$-LaZr$_{2}$H$_{24}$ respectively, which are close to those reported for established high-$T_c$ hydrides such as LaH$_{10}$ and
YH$_9$~\cite{somayazulu2019evidence,drozdov2019superconductivity,kong2019superconductivity}. 
By comparison, $\bar e/\mathrm{H}$ is found to be higher in  $I4/mmm$-La$_2$ZrH$_{12}$ (0.83), indicating a more ionic bonding character and reduced participation of hydrogen-derived states near the Fermi level. In contrast, the three identified systems, $Immm$-La$_3$H$_8$, $C2/m$-La$_4$H$_{17}$ and $I4/mmm$-LaZr$_2$H$_{12}$, exhibit lower H fraction at the Fermi level (see supplementary Fig.~7). Such reduced DOSs weaken the electron-phonon coupling and thus suppress superconducting properties~\cite{PhysRevB.110.214504}.\\

\noindent\textbf{Electron-phonon coupling and superconducting properties}\\
\noindent For the four selected compounds, the electron-phonon coupling strength was computed from first principles, and the resulting parameters were used to solve the isotropic Eliashberg equations numerically to evaluate $T_c$ and related superconducing properties as summerized in Table~1. A clear correlation is observed between the EPC constant $\lambda$ and the predicted $T_c$ across the four systems. The hydrogen-rich, high-symmetry phases $R3m$-Zr$_2$H$_{17}$ and $P\bar6m2$-LaZrH$_{18}$ exhibit the larger EPC constants (2.11 and 1.84, respectively), which give rise to high $T_c$ values of 178-209~K and 173-206~K at 300~GPa. Similarly, the $P6/mmm$-LaZr$_{2}$H$_{24}$ phase shows an even stronger EPC strength ($\lambda = 2.20$), yielding a predicted superconducting transition temperature of 202~K at a comparatively lower pressure of 200~GPa. Notably, a recent independent study~\cite{wang2026computational} reported a similar value of $T_c$ (209~K) for the same phase at 200~GPa, further supporting the robustness of the superconducting behavior in this hydrogen-rich ternary hydride.

In contrast, $I4/mmm$-La$_{2}$ZrH$_{12}$ exhibits substantially reduced EPC strength ($\lambda$ being 1.00), leading to markedly lower $T_c$ values of 75–97~K at 250~GPa. This systematic decrease in $T_c$ with decreasing $\lambda$ highlights the dominant role of the EPC strength in determining the superconducting performance of these hydrides at different pressures. The computed Eliashberg spectral functions (Fig.~\ref{Figure5}) further indicate that the large $\lambda$ values
in $R3m$-Zr$_2$H$_{17}$, $P\bar6m2$-LaZrH$_{18}$ and $P6/mmm$-LaZr$_{2}$H$_{24}$ are primarily driven by the strong coupling to high-frequency hydrogen-derived phonon modes, consistent with their dense hydrogen frameworks and enhanced hydrogen fraction near the Fermi level as discussed above. Overall, these results demonstrate that the strong EPC characterized by a large $\lambda$ in combination with favorable structural and electronic features is essential for achieving high-$T_c$ superconductivity in compressed hydrides.

\section*{Discussion}
Recent advances in data-driven approaches have significantly accelerated the discovery of high-temperature superconducting hydrides due to their computational efficiency. Here, we employed a previously established machine learning (ML) model~\cite{Zhao2024} to assess the superconducting potential of the hydrides identified through our structure searches.
The ML model is based on the Random Forest algorithm, offering a favorable balance between robustness and interpretability. It utilizes a minimal set of four physically meaningful descriptors: the standard deviation of valence electrons, the mean covalent radius, the range of Mendeleev numbers of constituent elements, and the hydrogen fraction of the total density of states at the Fermi level~\cite{Zhao2024}. 
When benchmarked against known superconducting hydrides, this ML model demonstrates competitive performance as compared to other data-driven approaches~\cite{belli2021strong, hutcheon2020predicting, wines2024data}, achieving a mean absolute error (MAE) of approximately 24 K and a root-mean-square error (RMSE) of about 34 K~\cite{Zhao2024}. 
We applied this ML model to all the hydride structures obtained from our USPEX searches and obtained reasonable agreement with our first-principles calculations for the 
four compounds considered in this work (Table 1, Supplementary Table 3).

We further compared our ML predictions to those obtained using the descriptor-based empirical model proposed by Belli et al.~\cite{belli2021strong}, which relates $T_c$ to hydrogen network connectivity, hydrogen fraction, and the hydrogen-projected density of states at the Fermi level. 
Overall, the Belli’s model yields reasonable $T_c$ values for the four considered compounds, as compared to the first-principles predictions (Table 1). While moderate deviations are  observed, they remain within the error bar of the model (30-50 K)~\cite{belli2021strong}. In contrast, the ML model by Zhao et al.~\cite{Zhao2024} was trained directly on $T_c$ data across a broader range of hydrides, enabling to capture nonlinear relationships between physically motivated descriptors and superconducting properties.
For example, the DFT-derived $T_c$ for $Fm\bar{3}m$-LaH$_{10}$ at 200 GPa is 271 K~\cite{kruglov2020superconductivity}, which agrees well with the ML-predicted value of 261 K. Similarly, for $P4/nmm$-LaH$_{11}$, the DFT-predicted $T_c$ at 150 GPa is 133 K~\cite{kruglov2020superconductivity}, in reasonable agreement with the ML result of 169.4 K (Supplementary Table 3).
In the case of metastable $C2/m$-LaH$_{23}$, DFT predicts $T_c$ = 101 K at 200 GPa~\cite{shutov2024ternary}, while the ML model yields a higher value of 171 K (Supplementary Table 3).
Applying the same ML model to the full set of thermodynamically and dynamically stable La-Zr-H compounds predicted in this work reveals that $P4/nmm$-ZrH$_{14}$ exhibits a $T_c$ of 128.2 K at 200 GPa. This is consistent with its high hydrogen contribution to the total DOS at the Fermi energy (0.38) and favorable $\bar{e}$/H ratio (0.28) (Supplementary Table 3).

To conclude, we have explored the La-Zr-H ternary hydride system under high-pressure conditions using the evolutionary structure search approach and first-principles calculations. 
Our analysis of thermodynamic and dynamic stabilities identified several novel hydride phases, 
including thermodynamically stable $R3m$-Zr$_{2}$H$_{17}$ and a low-enthalpy metastable phase $P\bar6m2$-LaZrH$_{18}$ at 300 GPa, which exhibit high superconducting $T_c$ values of 209 K and 206 K at 300 GPa, followed by another thermodynamically stable ternary phase, $P6/mmm$-LaZr$_{2}$H$_{24}$, with a predicted $T_c$ of 202~K at 200~GPa, consistent with a recent independent theoretical prediction for the same phase.
These three phases are featured by high crystal symmetry, densely-packed H cages, high phonon frequencies, high H fraction to the total DOS at the Fermi energy, and favorable value of $\bar{e}$/H, leading to strong electron-phonon coupling and resulting high $T_c$ values that calls for future experimental verification.
In contrast, $I4/mmm$-La$_{2}$ZrH$_{12}$, despite its high symmetry, exhibits comparatively lower $T_c$ values due to its higher $\bar{e}/\mathrm{H}$ ratio and reduced hydrogen content.
This work provides valuable insights for the design of high-$T_c$ hydrogen-rich superconducting materials with balanced structural and electronic characteristics.

\section*{Methods}
\noindent\textbf{Structure search and first-principles calculations}\\
Systematic structure searches were conducted to explore thermodynamically stable and metastable phases in the La-Zr-H system
using the evolutionary algorithm USPEX~\cite{lyakhov2013new, oganov2006crystal}. 
To efficiently explore the ternary system, a seed-based evolutionary structure prediction strategy was adopted. 
Specifically, variable-composition structure searches were initially performed at 0 GPa and 200 GPa for the La-H and Zr-H binary systems, which generated a diverse set of stable and low-energy binary hydride structures. The identified binary compounds were then employed as seeds in the evolutionary search for the ternary La-Zr-H system at pressures ranging from 150 to 300 GPa, with an interval of 50 GPa. 
This hierarchical approach improved the sampling efficiency and guided the evolutionary algorithm toward chemically meaningful regions of the La-Zr-H compositional space.
The convex hulls were all corrected for the ZPE,
which was obtained within the harmonic approximation. The structural optimizations and calculations of total energy and electronic properties were performed using the Vienna {\it ab initio} simulation package (VASP)~\cite{kresse1993ab, kresse1996efficient, kresse1996efficiency}.
The Perdew-Burke-Ernzerhof (PBE) exchange-correlation functional~\cite{Perdew1996_PBE} 
and projector-augmented wave (PAW) methods~\cite{KresseJoubert1999_PAW}
were used, which had been validated by previous high-pressure computations~\cite{Peng2017_HydrogenClathrate, Sun2020_LanthanidePolyhydrides, Oganov2009_IonicBoron}. A high plane-wave cutoff energy of 800 eV was adopted.
A $\Gamma$-centered $k$-point mesh with a spacing of $2\pi \times 0.016~\text{\AA}^{-1}$ was used for sampling the Brillouin zone.\\

\noindent\textbf{Phonon and EPC calculations}\\
Phonon dispersion relations and phonon densities of states of ternary La-Zr-H hydrides were computed using the finite-displacement method implemented in Phonopy ~\cite{Togo2015_phonon}, where interatomic force constants were extracted from density functional theory calculations using 2×2×2 supercells. Force calculations were performed using an energy cutoff of 800 eV and a denser $k$-point sampling ($2\pi \times 0.055\ \text{\AA}^{-1}$). The EPC's were calculated using the Quantum ESPRESSO (QE) package~\cite{Giannozzi2009_QE,PONCE2016116}. All calculations employed the PBE functional~\cite{Perdew1996_PBE} and Hartwigsen-Goedecker-Hutter (HGH) norm-conserving pseudopotentials~\cite{Allen1975} for La, Zr, and H atoms, with a plane-wave energy cutoff of 80 Ry. $T_c$ was calculated using the Allen-Dynes modified McMillan equation:
\begin{equation}
T_c = f_1 f_2 \frac{\omega_{\text{log}}}{1.2} \exp\left( \frac{-1.04(1 + \lambda)}{\lambda - \mu^* \left( 1 + 0.62 \lambda \right)} \right),
\end{equation}
which incorporates the strong coupling and shape correction factors, represented by the multipliers
\begin{equation}
f_1 = \left[ 1 + \left( \frac{\lambda}{2.46 \left( 1 + 3.8 \mu^* \right)} \right)^{\frac{3}{2}} \right]^{\frac{1}{3}}, 
\end{equation} 
\begin{equation}
f_2 = 1 + \frac{\left( \frac{\bar{\omega_2}}{\omega_{\text{log}}} - 1 \right) \lambda^2}{\lambda^2 + \left[ 1.82 \left( 1 + 6.3 \mu^* \right) \frac{\bar{\omega_2}}{\omega_{\text{log}}} \right]^2}, 
\end{equation} 
where \( \mu^{*} \) represents the Coulomb pseudopotential, typically ranging between 0.10 and 0.15; the parameters \( \lambda \), \( \bar\omega_2 \), and \( \omega_{\text{log}} \) denote the electron-phonon coupling constant, second moment of the normalized weight function, and logarithmic average frequency, respectively, and are defined as \begin{equation}
\lambda = \int_0^{\omega_{\text{max}}} \frac{2 \alpha^2 F(\omega)}{\omega} \, d\omega,
\end{equation}
\begin{equation}
\omega_{\text{log}} = \exp\left[ \frac{2}{\lambda} \int_0^{\omega_{\text{max}}} \frac{d\omega}{\omega} \alpha^2 F(\omega) \log(\omega) \right], 
\end{equation}
\begin{equation}
\bar\omega_2 = \sqrt{\frac{1}{\lambda} \int_0^{\omega_{\text{max}}} \left[ \frac{2 \alpha^2 F(\omega)}{\omega} \right] \omega^2 \, d\omega}, 
\end{equation}
where \( \alpha^2 F(\omega) \) is the Eliashberg spectral function calculated by
\begin{equation}
\alpha^2 F(\omega) = \frac{1}{2} \sum_v \int_{\text{BZ}} \frac{d\mathbf{q}}{\Omega_{\text{BZ}}} \lambda_{\mathbf{q}v} \, \omega_{\mathbf{q}v} \, \delta(\omega - \omega_{\mathbf{q}v}).
\end{equation}
Here, \( \Omega_{\text{BZ}} \) is volume of the Brillouin zone ,
\( \omega_{\mathbf{q}v} \) are the phonon frequencies,
and $\lambda_{\mathbf{q}v}$ are the mode-resolved electron-phonon couplings. The source code implementing this algorithm is publicly available on GitHub~\cite{shutov_a2f_2023}.

Additionally, the numerical solution of the isotropic Eliashberg equations was obtained using the iterative scheme propsed by Allen and Dynes~\cite{Allen1975,Migdal1958}.
We employed semi-empirical formulas~\cite{Carbotte1990} to calculate the zero-temperature superconducting gap \( \Delta(0) \) and thermodynamic critical magnetic field $H_c(0)$
\begin{equation}
\frac{2\Delta(0)}{k_B T_c} = 3.53 \left[ 1 + 12.5 \left( \frac{T_c}{\omega_{\log}} \right)^2 \ln \left( \frac{\omega_{\log}}{2T_c} \right) \right],
\end{equation}
\begin{equation}
\frac{\gamma T_c^2}{H_c^2(0)}
=
0.168 \left[
1 - 12.2 \left( \frac{T_c}{\omega_{\log}} \right)^2
\ln \left( \frac{\omega_{\log}}{3T_c} \right)
\right]
\tag{9}
\end{equation}
where $k_B$ is the Boltzmann constant A and 
$\gamma = \frac{2}{3}\pi^2 k_B^2 N(0)(1+\lambda)$ is the Sommerfeld coefficient. 
The values of $T_c$ from the solution of Eliashberg equations were used in these formulas.

To ensure consistency between the semi-empirical Allen–Dynes estimates and the direct numerical solution of the isotropic Eliashberg equations, the Coulomb pseudopotential parameter $(\text{\textmu}^*)$ was rescaled. While standard empirical values of $\text{\textmu}^{*}$ (typically 0.10–0.15) are optimized for the Allen–Dynes formula, the numerical solution of the Eliashberg equations on the imaginary axis requires an explicit cutoff defined by the maximum Matsubara frequency included in the discretization grid.

Following the renormalization procedure established by Allen and Dynes ~\cite{Pellegrini2024AbInitioSC, AllenDynes1974EliashbergCode}, the effective Coulomb pseudopotential for the Eliashberg equation $(\mu^{*}_{Eliashberg})$ was derived from the input McMillan value $(\mu^{*}_{McMillan})$ using the relation:
\begin{equation}
\frac{1}{\mu^{*}_{Eliashberg}}
=
\frac{1}{\mu^{*}_{McMillan}}
+
\ln \left(
\frac{\omega_{ph}}{\omega_{\mathit{max}}^{Matsub}}
\right),
\tag{10}
\end{equation}
where $\omega_{ph}$ denotes the maximum phonon frequency. 
The cutoff frequency $\omega_{\max}^{Matsub}$ corresponds to the highest Matsubara frequency retained in the calculation, which is defined  as: 
\begin{equation}
\hbar \omega_{\mathit{max}}^{\mathit{Matsub}}
=
2\pi N k_B T_c^{AD}.
\tag{11}
\end{equation}
Here, \( \mathit{k}\mathrm{_B} \) is the Boltzmann constant, $T_c^{AD}$ is the critical temperature estimated via the Allen–Dynes equation, and $N$ represents the matrix dimension in the numerical solution of the Eliashberg equations and is equal to 24.


\section*{Data Availability}
The data supporting the findings of this study are included within the manuscript or can be obtained from the corresponding author upon reasonable request.

\section*{Code Availability}
VASP can be acquired from the VASP Software GmbH (see \href{https://www.vasp.at/}{www.vasp.at});
Phonopy is available at \href{https://phonopy.github.io/phonopy/}{phonopy.github.io/phonopy};
Quantum ESPRESSO is available at \href{https://www.quantum-espresso.org}{www.quantum-espresso.org}.

\bibliography{Reference-MS}
\bibliographystyle{naturemag}
\expandafter\ifx\csname url\endcsname\relax
  \def\url#1{\texttt{#1}}\fi
\expandafter\ifx\csname urlprefix\endcsname\relax\def\urlprefix{URL }\fi
\providecommand{\bibinfo}[2]{#2}
\providecommand{\eprint}[2][]{\url{#2}}


\section*{Acknowledgements}
This work is supported by
the National Natural Science Foundation of China (Grants No.~52422112 and No.~52188101),
the Strategic Priority Research Program of the Chinese Academy of Sciences (XDA041040402),
the Liaoning Province Science and Technology Major Project (2024JH1/11700032, 2023021207-JH26/103 and RC230958),
the National Key R{\&}D Program of China 2021YFB3501503,
the Special Projects of the Central Government in Guidance of Local Science and Technology Development (2024010859-JH6/1006), and the Russian Science Foundation (Grant No. 19-72-30043). This research partly used computational resources of the HPC SEC shared research facilities at UNN.

\section*{Author Contributions}
X.-Q.C. and P.L. conceived the project and designed the research.
I.S., M.A.G., and J.Z. performed the calculations.
P.L., Y.S., M.A.G., A.R.O., and X.-Q.C supervised the project.
E.B., T.Y., X.L. and Y.Z. participated in discussions.
I.S. and P.L. wrote the manuscript with inputs from other authors.
All authors commented on the manuscript.
I.S., M.A.G. and J.Z. contributed equally to this work.

\section*{Competing Interests}
The authors declare no competing interests.

\end{document}


\captionsetup[figure]{labelfont={bf},labelformat={default},labelsep=period,name={Supplementary Fig.}}
\captionsetup[table]{labelfont={bf},labelformat={default},labelsep=period,name={Supplementary Table}}

		\maketitle
		\vspace{-5mm}
		\begin{center}
			\begin{minipage}{1\textwidth}
				\begin{center}
					\textit{
						\textsuperscript{1} Shenyang National Laboratory for Materials Science, Institute of Metal Research, Chinese Academy of Sciences, Shenyang 110016, China
						\\\textsuperscript{2} School of Materials Science and Engineering, University of Science and Technology of China, Shenyang 110016, China
						\\\textsuperscript{3} Skolkovo Institute of Science and Technology, Bolshoy Boulevard 30, bld. 1, 121205, Moscow, Russia
						\\\textsuperscript{4} Department of Computer Science, Xinzhou Normal University, Xinzhou 034000, China
						\vspace{5mm}
						\\{$\dagger$} Corresponding to: ptliu@imr.ac.cn, xingqiu.chen@imr.ac.cn
						\\{$\star$} These authors contribute equally.
						\vspace{5mm}
					}
				\end{center}
			\end{minipage}
		\end{center}
		
		\setlength\parindent{13pt}

	\newpage
	\clearpage

\section*{Supplementary Figures}
\vspace{4mm}

\begin{figure*}[ht!]
\centering
\includegraphics[width=\textwidth]{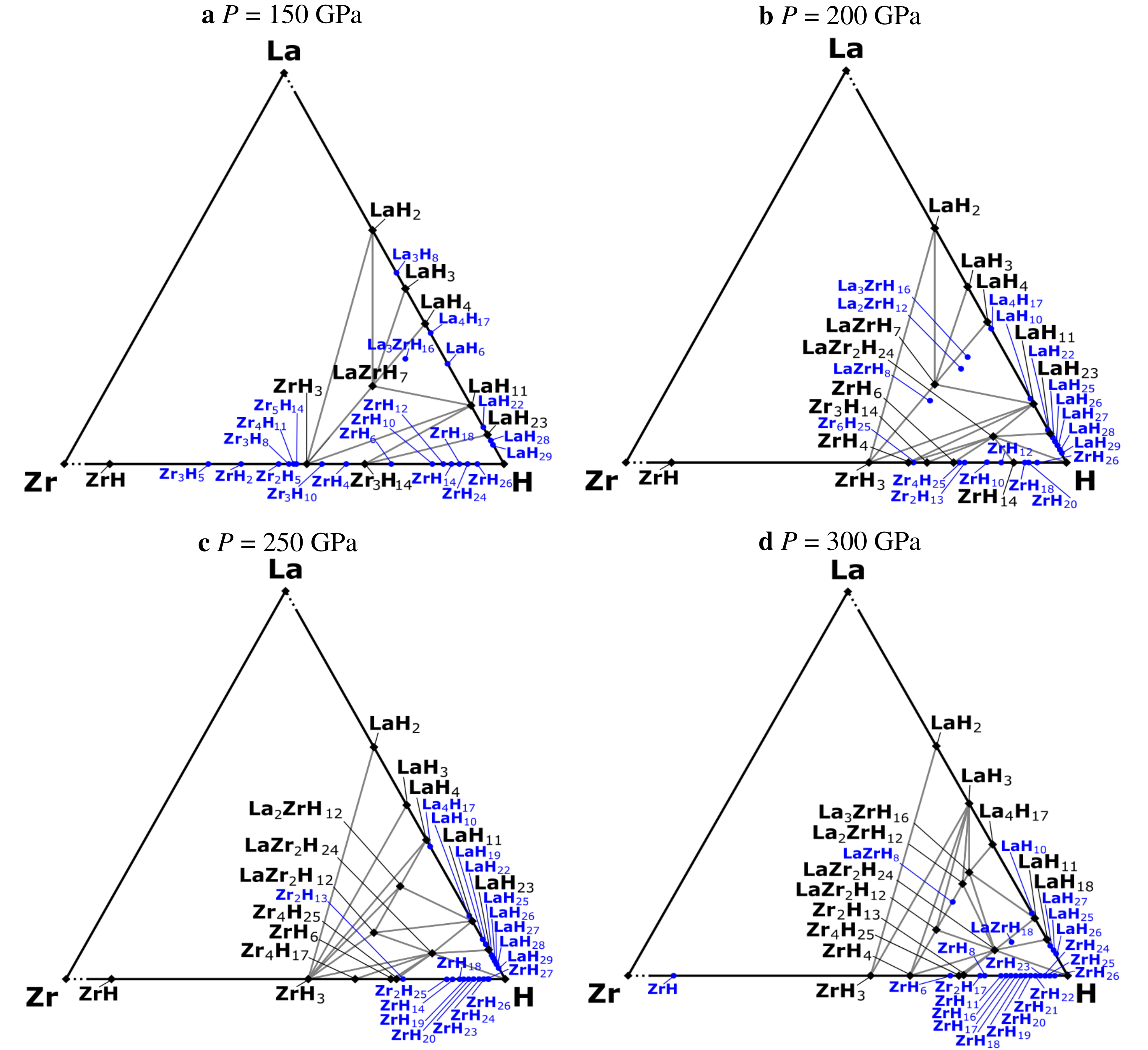} 
		\caption{\textbf{The non-ZPE-corrected convex hulls and phase diagrams for the La-Zr-H system at} \textbf {a}. 150 GPa \textbf{b}. 200 GPa \textbf{c}. 250 GPa and \textbf{d}. 300 GPa. Black diamonds represent thermodynamically stable phases, while blue circles denote metastable structures with formation enthalpies within 30 meV/atom above the convex hull.
}
		\label{fig:FigS1}
	\end{figure*}

\begin{figure*}[ht!]
\centering
\includegraphics[width=\textwidth]{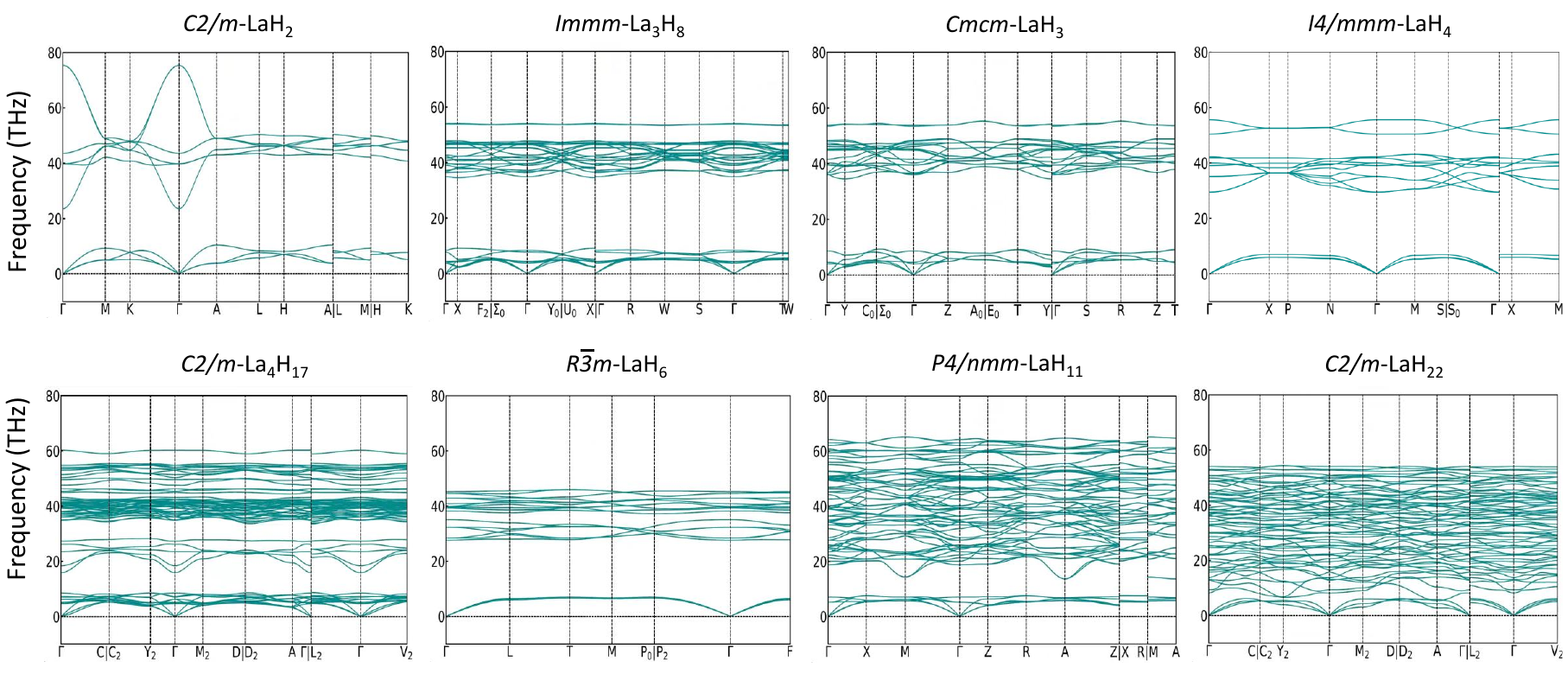} 
\includegraphics[width=\textwidth]{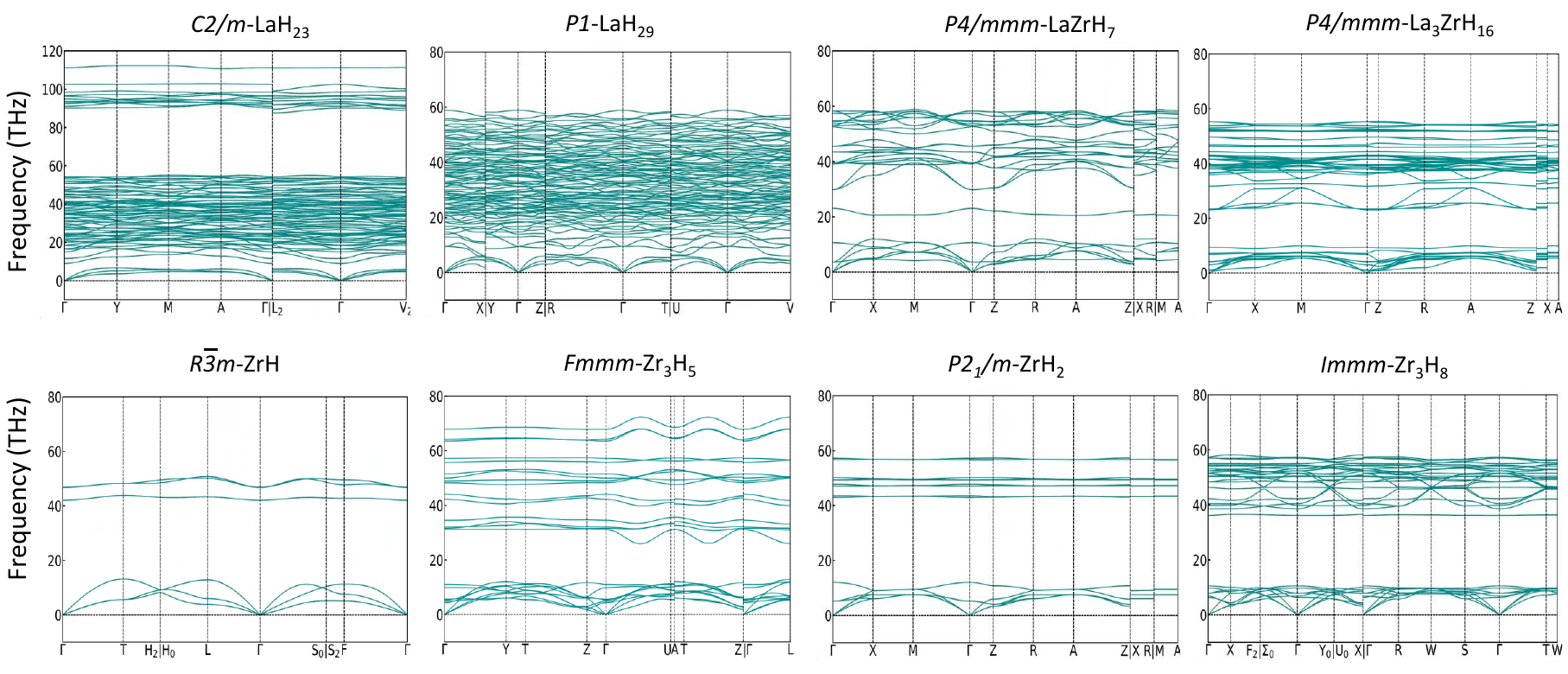} 
\includegraphics[width=\textwidth]{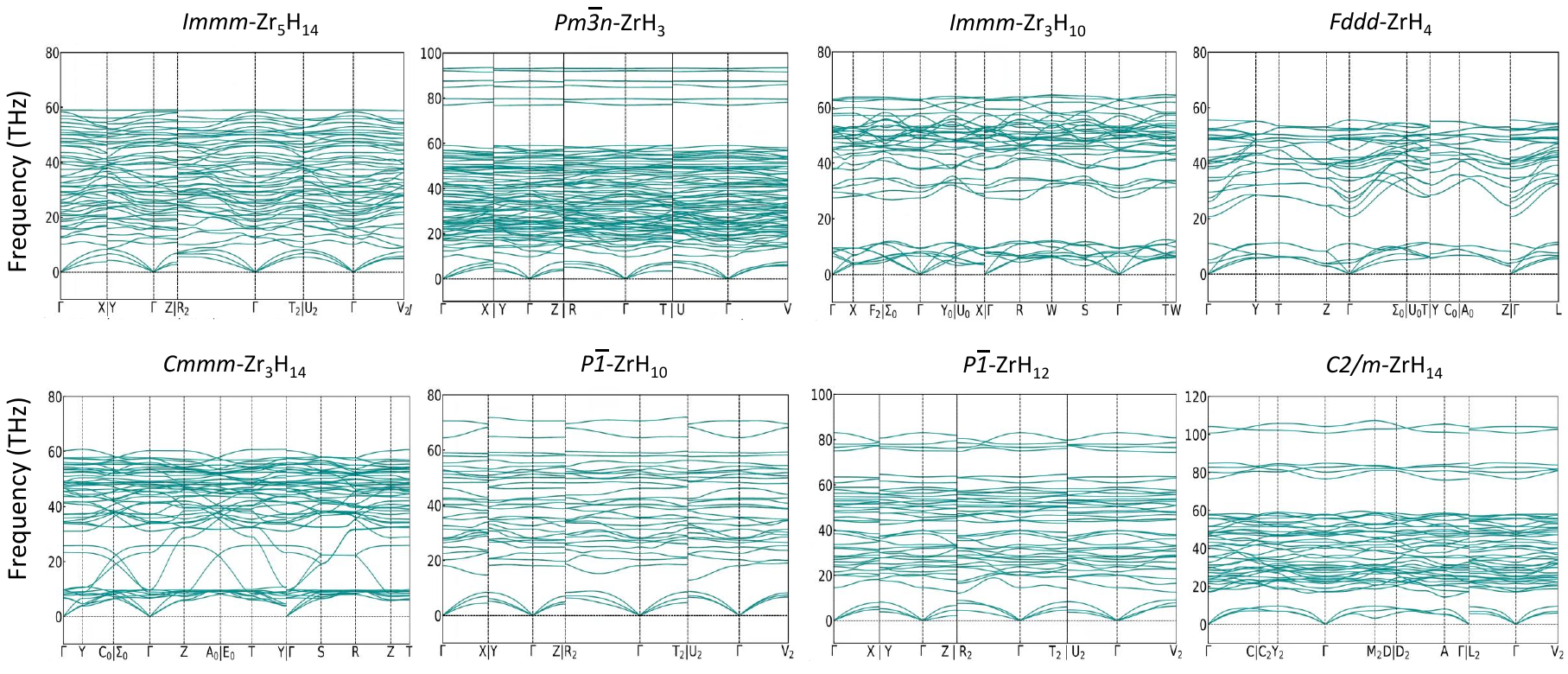}
\end{figure*}

\clearpage

\begin{figure*}[ht!]
\centering
\includegraphics[width=\textwidth]{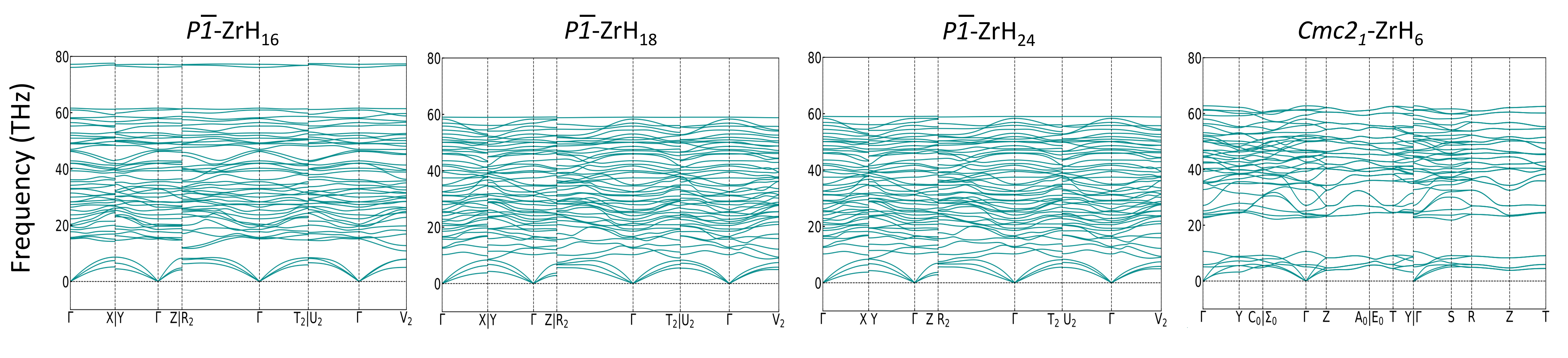}
\caption{\textbf{Phonon dispersion relationships of dynamically stable structures in the La–Zr–H system at 150 GPa.}}
\label{fig:FigS1}
\end{figure*}

\begin{figure*}[ht!]
\centering
\includegraphics[width=\textwidth]{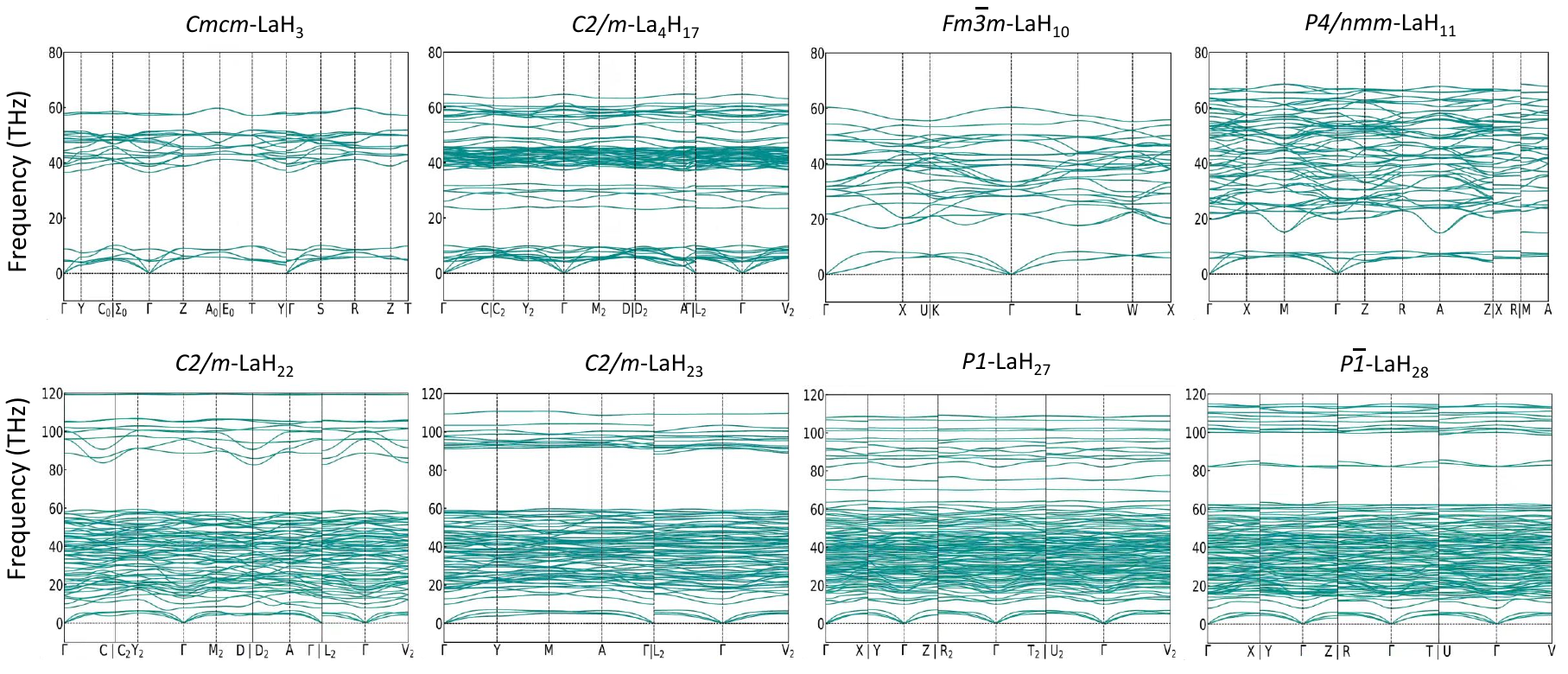} 
\includegraphics[width=\textwidth]{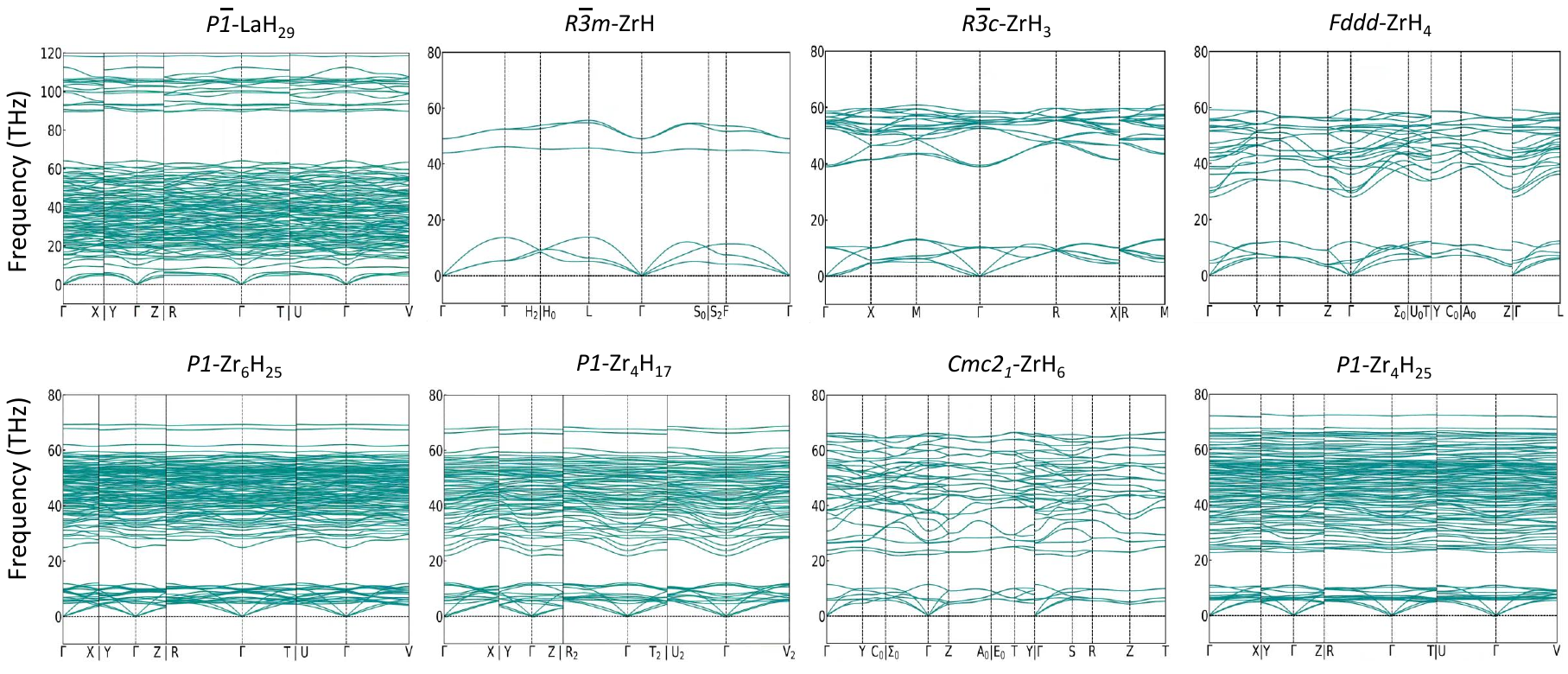} 
\includegraphics[width=\textwidth]{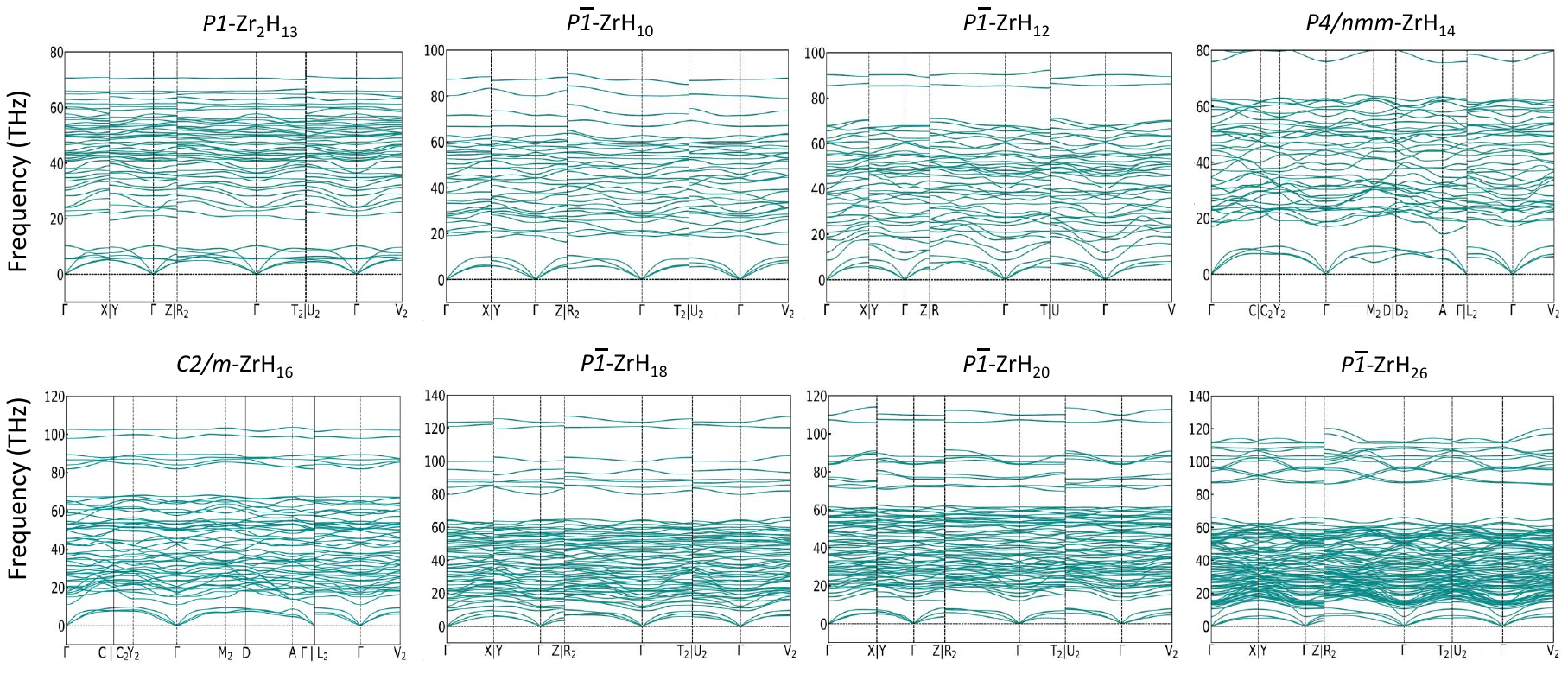}
\end{figure*}

\clearpage

\begin{figure*}[ht!]
\centering
\includegraphics[width=\textwidth]{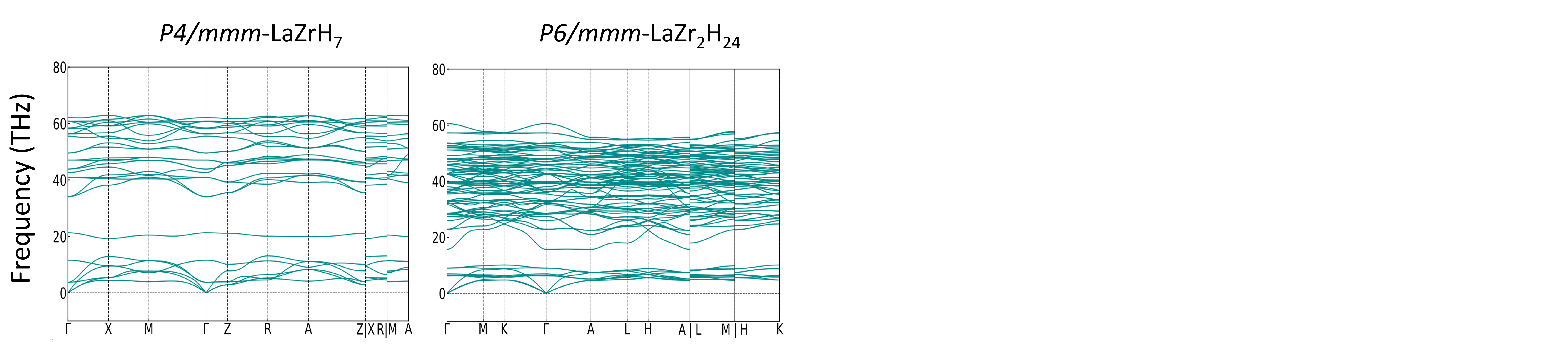}
\caption{\textbf{Phonon dispersion relationships of dynamically stable structures in the La–Zr–H system at 200 GPa.}}
\label{fig:FigS1}
\end{figure*}

\begin{figure*}[ht!]
\centering
\includegraphics[width=\textwidth]{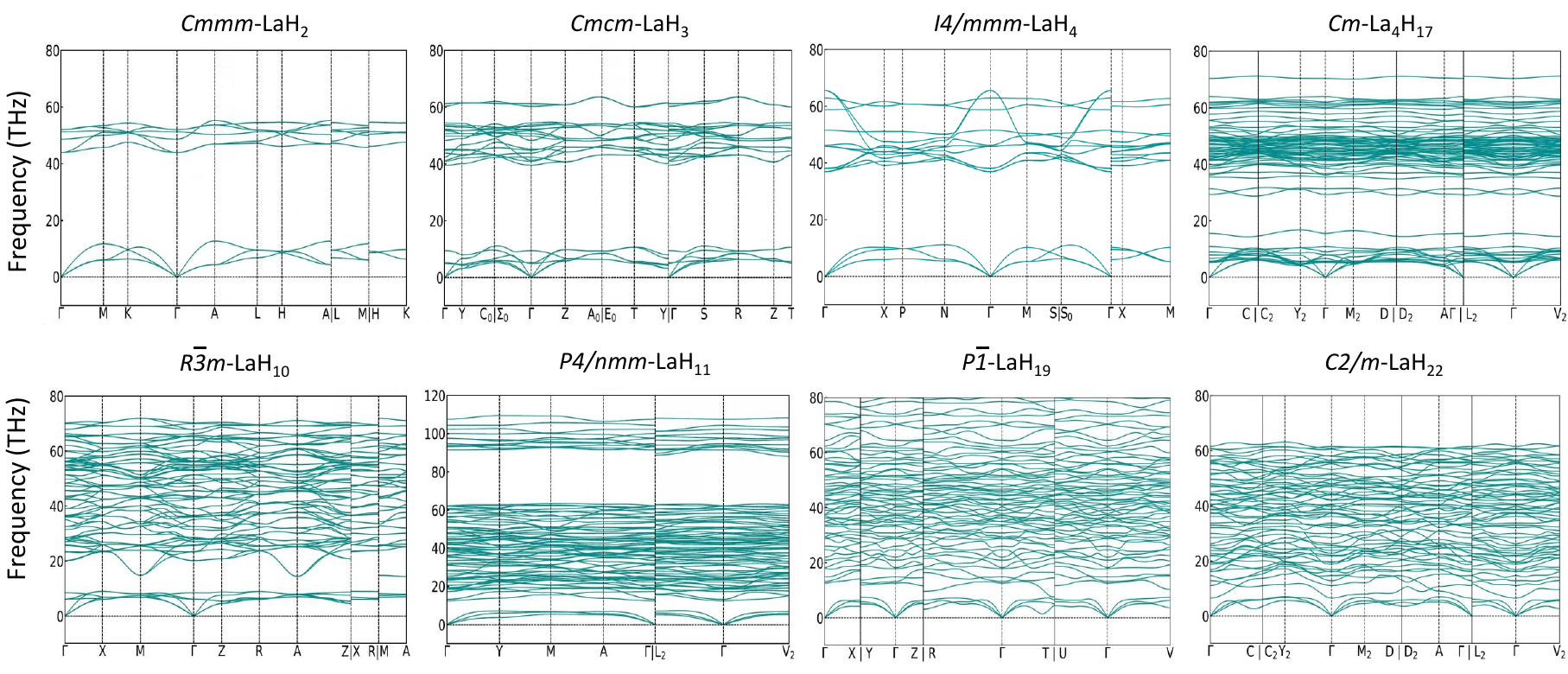} 
\includegraphics[width=\textwidth]{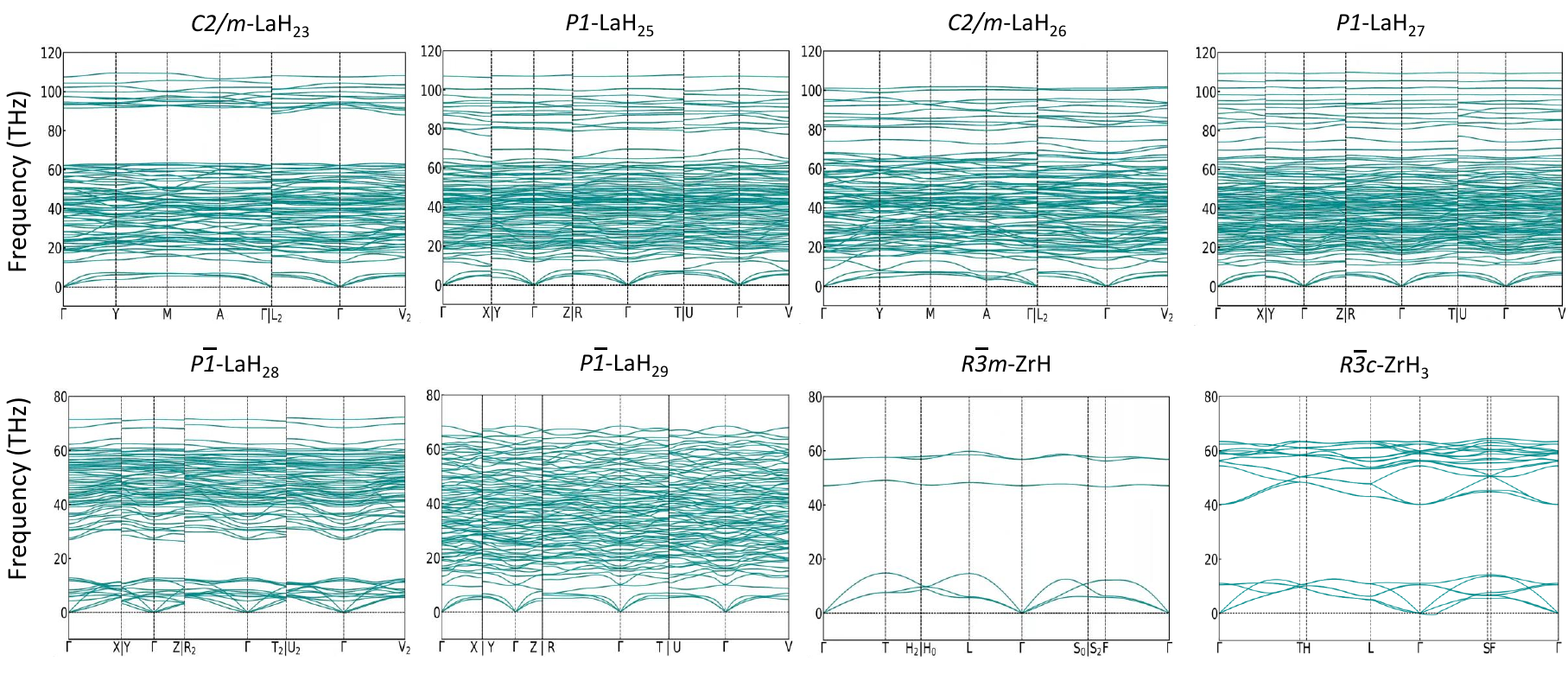} 
\includegraphics[width=\textwidth]{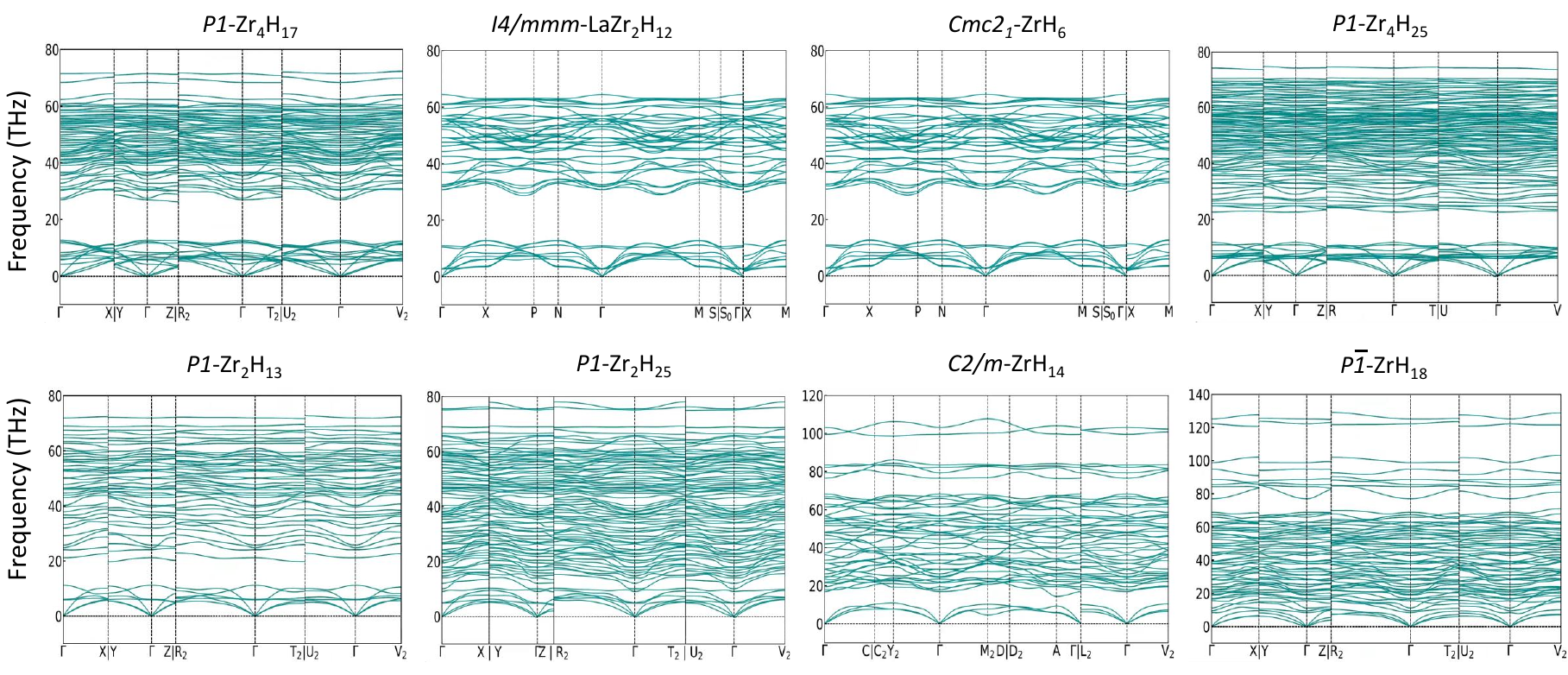}
\end{figure*}

\clearpage

\begin{figure*}[ht!]
\centering
\includegraphics[width=\textwidth]{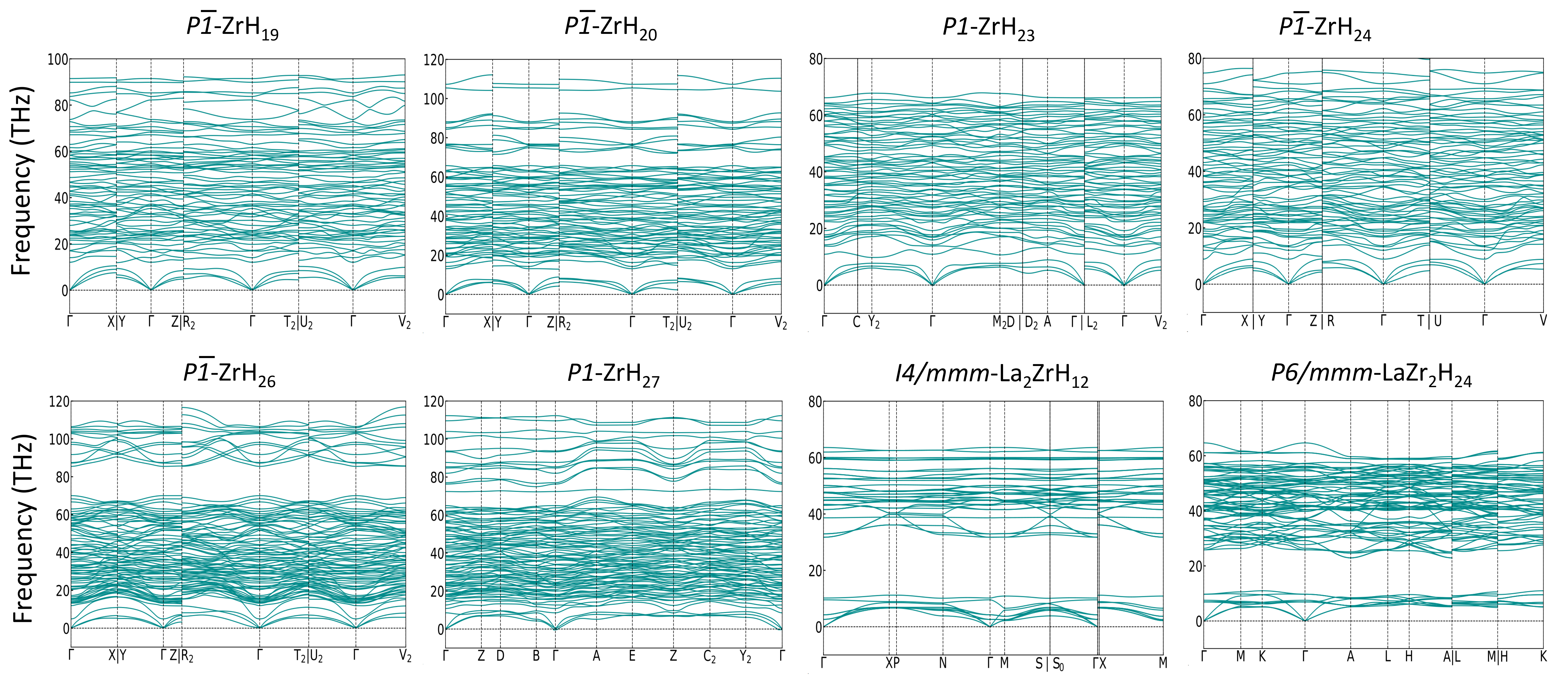}
\caption{\textbf{Phonon dispersion relationships of dynamically stable structures in the La–Zr–H system at 250 GPa.}}
\label{fig:FigS1}
\end{figure*}

\begin{figure*}[ht!]
\centering
\includegraphics[width=\textwidth]{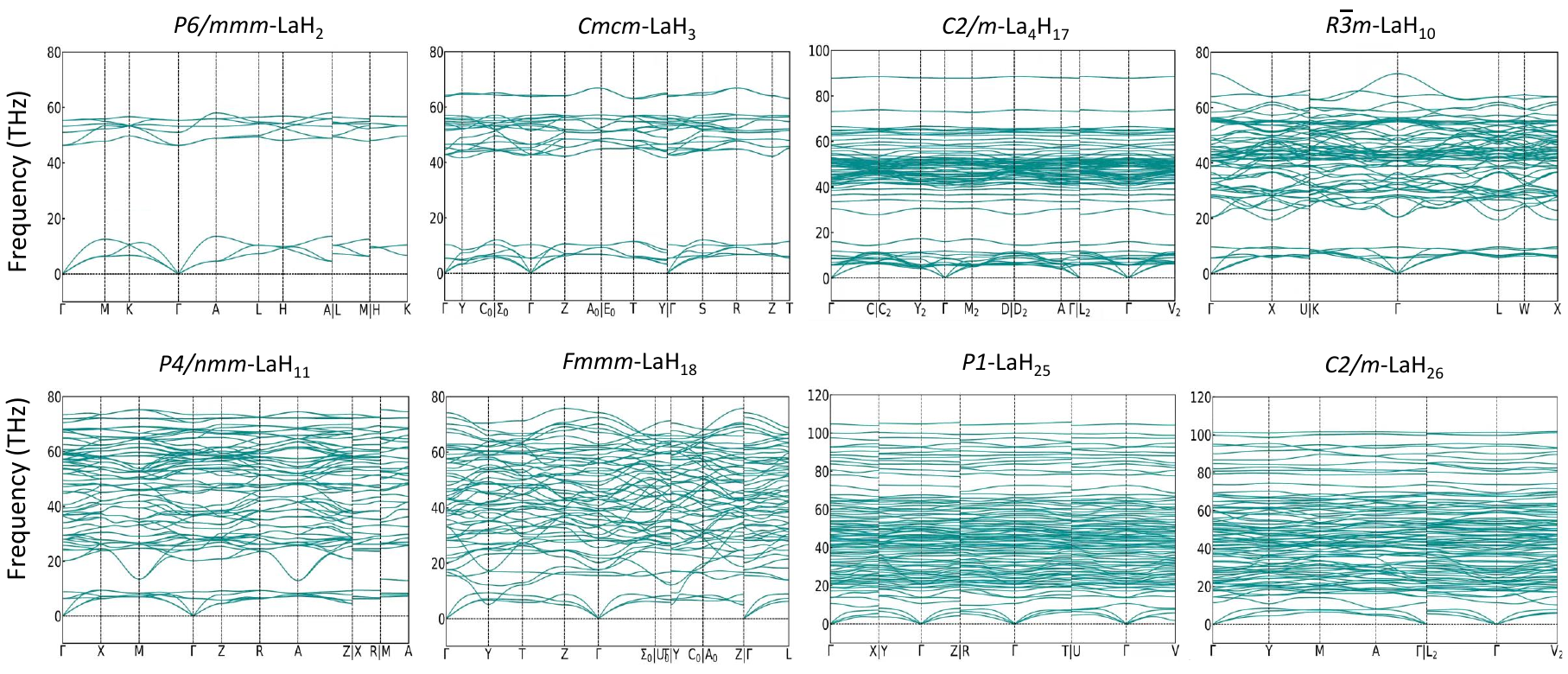} 
\includegraphics[width=\textwidth]{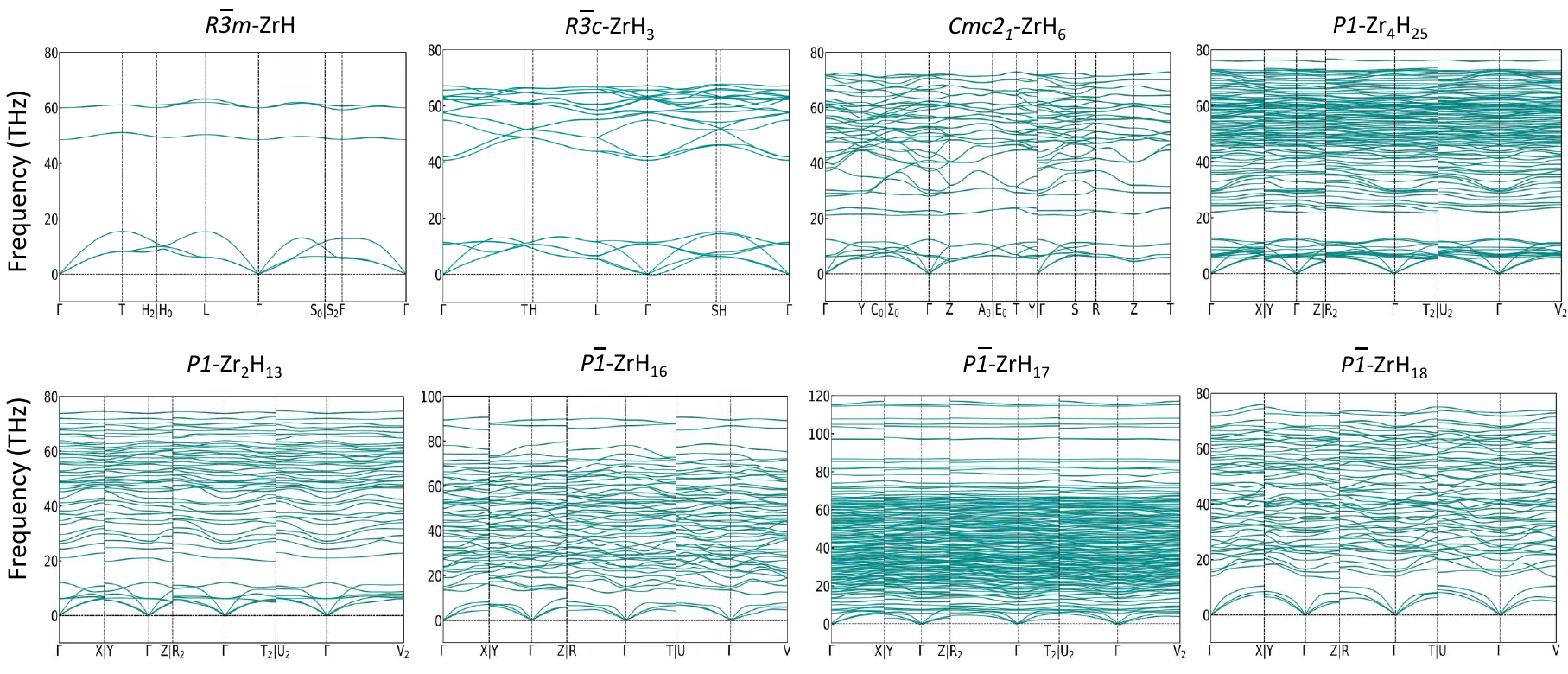} 
\includegraphics[width=\textwidth]{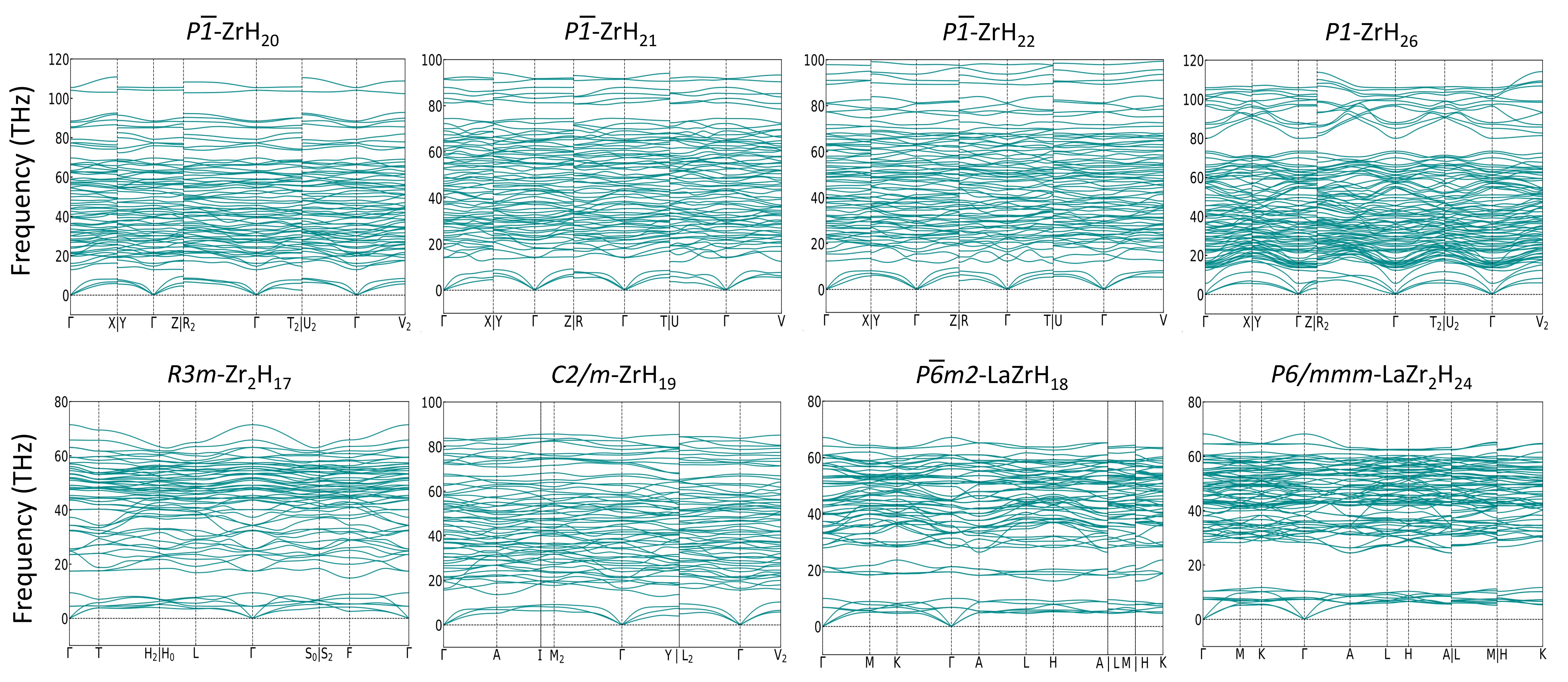}
\caption{\textbf{Phonon dispersion relationships of dynamically stable structures in the La–Zr–H system at 300 GPa.}
}
		\label{fig:FigS2}
\end{figure*}  

\clearpage
	\begin{figure*}[ht!]
\centering
\includegraphics[width=\textwidth]{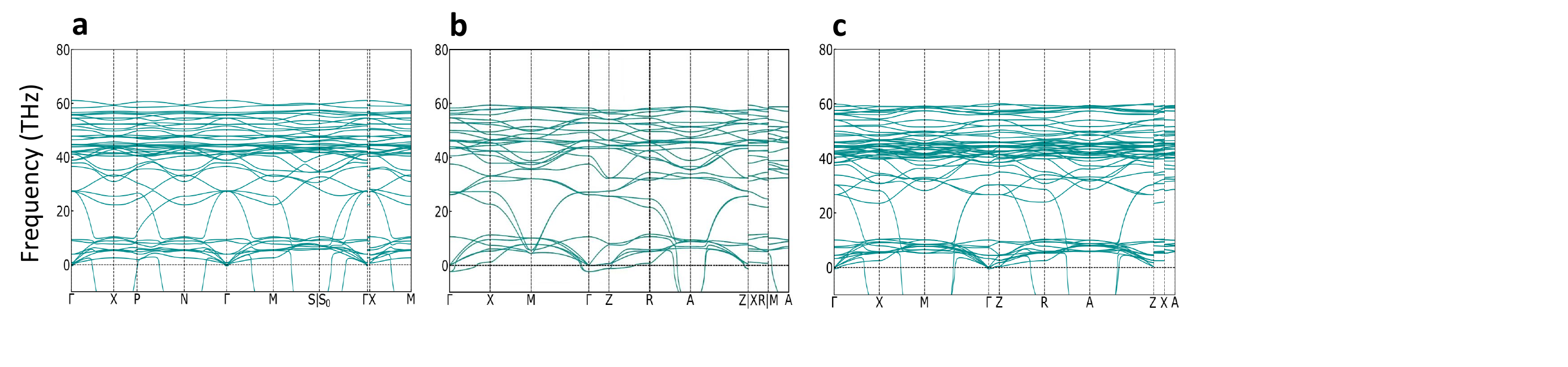} 
		\caption{\textbf{Phonon dispersion relationships of identified thermodynamically stale but dynamically unstable ternary systems at 200 GPa.}
\textbf{a}. $I4/mmm$-La$_{2}$ZrH$_{12}$
\textbf{b}. $P4/mmm$-LaZrH$_{8}$.
\textbf{c}. $Pmm2$-La$_{3}$ZrH$_{16}$.
}
		\label{fig:FigS5}
	\end{figure*}

\clearpage
\begin{figure*}[ht!]
\centering
\includegraphics[width=\textwidth]{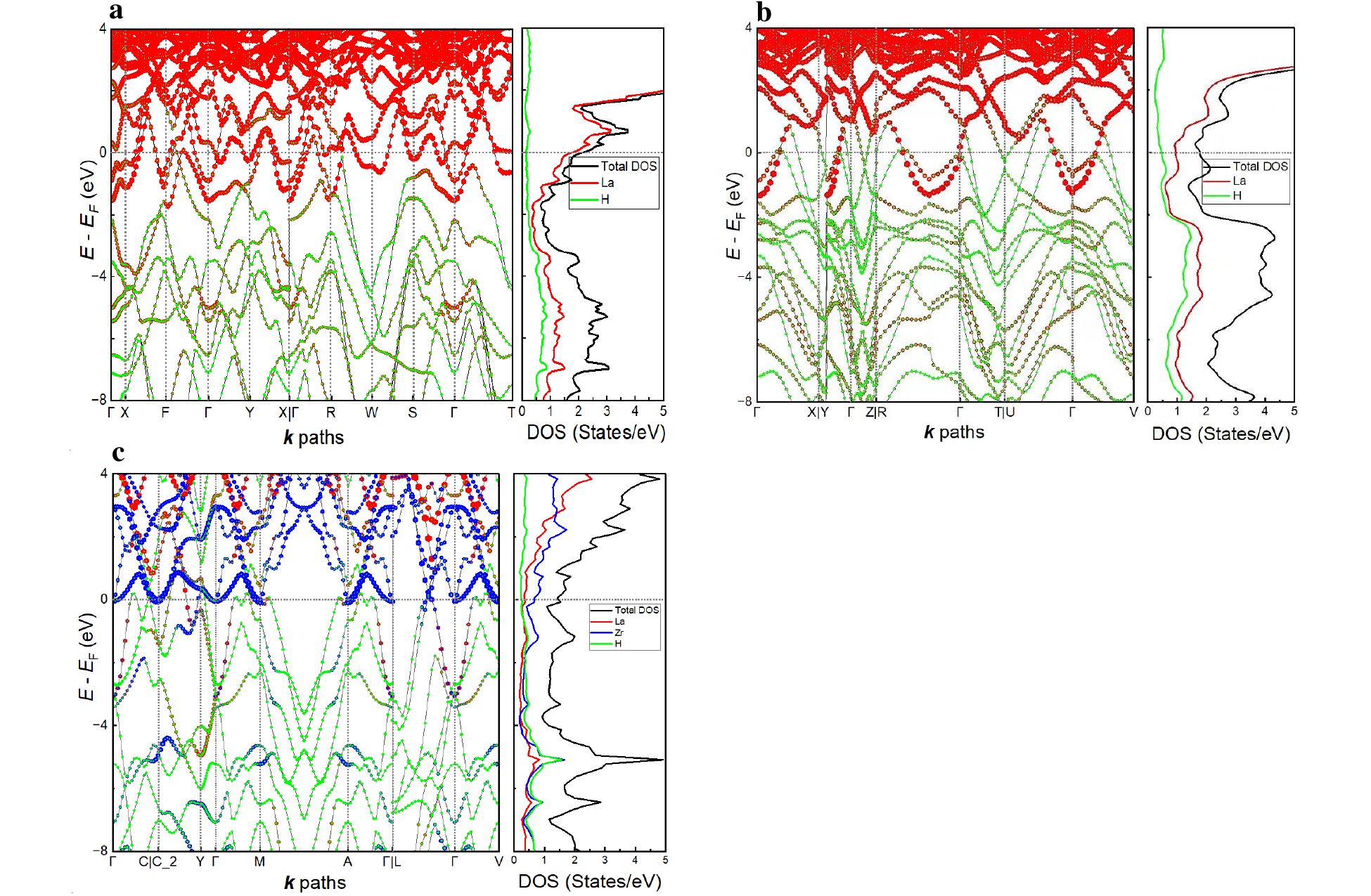} 
\caption{\textbf{Electronic band structures and DOSs.} 
\textbf {a}. $Immm$-La$_{3}$H$_{8}$ and
\textbf{b}. $C2/m$-La$_{4}$H$_{17}$ at 150 GPa.
\textbf{c}. $I4/mmm$-LaZr$_{2}$H$_{12}$ at 250 GPa.
		}
\label{fig:FigS6}
\end{figure*}

\clearpage
\section*{Supplementary Tables}
\vspace{4mm}

\setlength{\fboxrule}{1.1pt}   
\setlength{\fboxsep}{4pt}     
\setlength{\arrayrulewidth}{1.0pt}  

\begin{table*}[!htbp]
\centering
\label{tab:uspex-summary}
\renewcommand{\arraystretch}{1.3}

\fbox{%
\begin{minipage}{\textwidth}
\small
\textbf{Supplemntary Table 1.} Initial USPEX structure search in {{La}}-{{H}} and {{Zr}}–{{H}} binary hydrides with variable compositions as seeds for subsequent ternary structure search.
For each search, the first generation consists of 120 species, followed by subsequent generations containing 100 species each. New candidate structures were generated using heredity (40\%), random structure generator (30\%), and mutation (30\%).\\[4pt]
\noindent\rule{\textwidth}{1.5pt}  
\resizebox{\textwidth}{!}{%
\begin{tabular}{p{3.5cm} p{2.0cm} p{2.0cm} p{4.5cm} p{4.5cm}}
\toprule
\textbf{No. of atoms} & \textbf{Pressure} & \textbf{System} & \textbf{Stable phases} & \textbf{Metastable phases} \\
\midrule
Minimum: 6 \newline Maximum: 18 & 0 GPa & {{La}}-{{H}} &
$Fm\bar{3}m$-La$_{3}$H$_{6}$ \newline
$Immm$-La$_{3}$H$_{7}$ \newline
$Immm$-La$_{3}$H$_{8}$ \newline
$C2/m$-La$_{3}$H$_{9}$ &
$Fm\bar{3}m$-La$_{3}$H$_{8}$ \newline
$Pmmn$-La$_{2}$H$_{6}$ \\
\hline
Minimum: 4 \newline Maximum: 16 & 200 GPa & {{Zr}}–{{H}} &
$R\bar{3}m$-Zr$_{2}$H$_{2}$ \newline
$R\bar{3}c$-Zr$_{2}$H$_{6}$ \newline
$Cmc2_1$-Zr$_{2}$H$_{12}$ \newline
$I4/mmm$-Zr$_{2}$H$_{8}$ &
$R\bar{3}m$-Zr$_{5}$H$_{5}$ \newline
$R\bar{3}m$-Zr$_{5}$H$_{6}$ \newline
$Fmmm$-Zr$_{3}$H$_{12}$ \newline
$C2/m$-Zr$_{4}$H$_{14}$ \\
\hline
Minimum: 8 \newline Maximum: 25 & 200 GPa & {{La}}-{{H}} &
$I4/mmm$-La$_{2}$H$_{12}$ \newline
$C2/m$-La$_{3}$H$_{10}$ \newline
$P4/nmm$-La$_{2}$H$_{22}$ &
$C2/m$-La$_{4}$H$_{16}$ \newline
$R\bar{3}m$-La$_{2}$H$_{20}$ \\
\bottomrule
\end{tabular}
}
\end{minipage}
}
\end{table*}

\vspace{12pt}
\small

\begin{mdframed}[
linewidth=1.1pt,
innerleftmargin=4pt,
innerrightmargin=4pt,
innertopmargin=6pt,
innerbottommargin=6pt,
splittopskip=\topskip,
splitbottomskip=\topskip,
nobreak=false,
hidealllines=false,
leftline=true,
rightline=true,
topline=true,
bottomline=true
]

\centering
\textbf{Supplementary Table 2.}
\centering
Structural data of identified novel binary and ternary hydrides at different pressures.

\medskip
\noindent\rule{\textwidth}{1pt}

\vspace{-6pt}

\begin{longtable}{l c c c l}

\textbf{Structure} &
\textbf{Pressure (GPa)} &
\textbf{Lattice parameters} &
\textbf{Volume (\AA$^3$/atom)} &
\textbf{Atomic coordinates} \\
\hline
\endfirsthead

\hline
\endfoot

\bottomrule
\endlastfoot

\toprule
\textbf{Structure} &
\textbf{Pressure (GPa)} &
\textbf{Lattice parameters} &
\textbf{Volume (\AA$^3$/atom)} &
\textbf{Atomic coordinates} \\
\hline
\endhead

$C2/m$-La$_4$H$_{17}$ & 150 &
\begin{tabular}[t]{l}
$a=10.275$ \AA \\
$b=2.763$ \AA \\
$c=6.636$ \AA \\
$\beta=101.187^\circ$
\end{tabular}
& 4.400 &
\begin{tabular}[t]{lccc}
La1 & 0.124 & 0.500 & 0.621 \\
La2 & 0.132 & 0.000 & 0.130 \\
H1 & 0.037 & 0.000 & 0.386 \\
H2 & 0.040 & 0.500 & 0.872 \\
H3 & 0.060 & 0.000 & 0.817 \\
H4 & 0.061 & 0.500 & 0.307 \\
H5 & 0.182 & 0.500 & 0.935 \\
H6 & 0.189 & 0.000 & 0.438 \\
H7 & 0.209 & 0.000 & 0.856 \\
H8 & 0.214 & 0.500 & 0.367 \\
H9 & 0.000 & 0.500 & 0.000 \\
\end{tabular}
\\ \hline

$Cmmm$-Zr$_3$H$_{14}$ & 150 &
\begin{tabular}[t]{l}
$a=3.301$ \AA \\
$b=11.821$ \AA \\
$c=3.242$ \AA
\end{tabular}
& 3.720 &
\begin{tabular}[t]{lccc}
Zr1 & 0.000 & 0.351 & 0.500 \\
Zr2 & 0.000 & 0.000 & 0.000 \\
H1 & 0.000 & 0.212 & 0.260 \\
H2 & 0.239 & 0.137 & 0.000 \\
H3 & 0.000 & 0.074 & 0.500 \\
H4 & 0.000 & 0.264 & 0.000 \\
H5 & 0.000 & 0.500 & 0.257 \\
\end{tabular}
\\ \hline

$Immm$-La$_3$H$_8$ & 150 &
\begin{tabular}[t]{l}
$a=2.728$ \AA \\
$b=2.790$ \AA \\
$c=15.744$ \AA
\end{tabular}
& 5.447 &
\begin{tabular}[t]{lccc}
La1 & 0.000 & 0.500 & 0.159 \\
La2 & 0.000 & 0.000 & 0.000 \\
H1 & 0.000 & 0.000 & 0.250 \\
H2 & 0.000 & 0.000 & 0.448 \\
H3 & 0.000 & 0.500 & 0.293 \\
H4 & 0.000 & 0.500 & 0.399 \\
\end{tabular}
\\ \hline
\hline  

$P\bar{1}$-Zr$_2$H$_{13}$ & 250 &
\begin{tabular}[t]{l}
$a=3.022$ \AA \\
$b=3.094$ \AA \\
$c=5.008$ \AA \\
$\alpha=90.4773^\circ$ \\
$\beta=89.9765^\circ$ \\
$\gamma=118.0037^\circ$
\end{tabular}
& 2.756 &
\begin{tabular}[t]{lccc}
Zr1 & 0.547 & 0.799 & 0.216 \\
Zr2 & 0.892 & 0.495 & 0.736 \\
H1 & 0.416 & 0.985 & 0.913 \\
H2 & 0.068 & 0.817 & 0.427 \\
H3 & 0.214 & 0.156 & 0.212 \\
H4 & 0.528 & 0.856 & 0.618 \\
H5 & 0.554 & 0.295 & 0.421 \\
H6 & 0.144 & 0.969 & 0.020 \\
H7 & 0.870 & 0.442 & 0.113 \\
H8 & 0.373 & 0.464 & 0.929 \\
H9 & 0.856 & 0.988 & 0.915 \\
H10 & 0.938 & 0.991 & 0.524 \\
H11 & 0.225 & 0.147 & 0.719 \\
H12 & 0.378 & 0.413 & 0.516 \\
H13 & 0.017 & 0.338 & 0.401 \\
\end{tabular}
\\ \hline

$R3m$-Zr$_2$H$_{17}$ & 300 &
\begin{tabular}[t]{l}
$a=3.129$ \AA \\
$c=15.456$ \AA
\end{tabular}
& 2.299 &
\begin{tabular}[t]{lccc}
Zr1 & 0.000 & 0.000 & 0.034 \\
Zr2 & 0.000 & 0.000 & 0.532 \\
H1 & 0.027 & 0.514 & 0.095 \\
H2 & 0.030 & 0.515 & 0.968 \\
H3 & 0.033 & 0.516 & 0.593 \\
H4 & 0.179 & 0.359 & 0.806 \\
H5 & 0.000 & 0.000 & 0.211 \\
H6 & 0.000 & 0.000 & 0.357 \\
H7 & 0.000 & 0.000 & 0.657 \\
H8 & 0.000 & 0.000 & 0.728 \\
H9 & 0.000 & 0.000 & 0.847 \\
\end{tabular}
\\ \hline

$P6/mmm$-LaZr$_2$H$_{24}$ & 200 &
\begin{tabular}[t]{l}
$a=4.978$ \AA \\
$c=3.491$ \AA
\end{tabular}
& 2.775 &
\begin{tabular}[t]{lccc}
La1 & 0.000 & 0.000 & 0.000 \\
Zr1 & 0.333 & 0.667 & 0.500 \\
H1 & 0.000 & 0.379 & 0.208 \\
H2 & 0.000 & 0.221 & 0.500 \\
H3 & 0.237 & 0.475 & 0.000 \\
\end{tabular}
\\ \hline

$I4/mmm$-La$_2$ZrH$_{12}$ & 250 &
\begin{tabular}[t]{l}
$a=2.594$ \AA \\
$c=16.521$ \AA
\end{tabular}
& 3.706 &
\begin{tabular}[t]{lccc}
La1 & 0.000 & 0.000 & 0.336 \\
Zr1 & 0.000 & 0.000 & 0.000 \\
H1 & 0.000 & 0.500 & 0.074 \\
H2 & 0.000 & 0.000 & 0.110 \\
H3 & 0.000 & 0.000 & 0.214 \\
H4 & 0.000 & 0.000 & 0.458 \\
H5 & 0.000 & 0.500 & 0.250 \\
\end{tabular}
\\ \hline

$I4/mmm$-LaZr$_2$H$_{12}$ & 250 &
\begin{tabular}[t]{l}
$a=2.610$ \AA \\
$c=15.572$ \AA
\end{tabular}
& 3.537 &
\begin{tabular}[t]{lccc}
La1 & 0.000 & 0.000 & 0.500 \\
Zr1 & 0.000 & 0.000 & 0.169 \\
H1 & 0.000 & 0.500 & 0.092 \\
H2 & 0.000 & 0.000 & 0.051 \\
H3 & 0.000 & 0.000 & 0.285 \\
H4 & 0.000 & 0.000 & 0.369 \\
H5 & 0.000 & 0.500 & 0.250 \\
\end{tabular}
\\ \hline

$Pm2$-LaZrH$_{18}$ & 300 &
\begin{tabular}[t]{l}
$a=3.342$ \AA \\
$c=4.950$ \AA
\end{tabular}
& 2.393 &
\begin{tabular}[t]{lccc}
La1 & 0.000 & 0.000 & 0.500 \\
Zr1 & 0.333 & 0.667 & 0.000 \\
H1 & 0.358 & 0.179 & 0.173 \\
H2 & 0.514 & 0.029 & 0.303 \\
H3 & 0.000 & 0.000 & 0.109 \\
H4 & 0.333 & 0.667 & 0.599 \\
H5 & 0.667 & 0.333 & 0.000 \\
H6 & 0.667 & 0.333 & 0.500 \\
\end{tabular}
\\

\end{longtable}

\end{mdframed}

\setlength{\fboxrule}{1.1pt}   
\setlength{\fboxsep}{4pt}     
\setlength{\arrayrulewidth}{1.0pt}  

\begin{table*}[t]
\centering
\label{tab:tc_properties_fullpage}
\renewcommand{\arraystretch}{1.3}

\fbox{%
\begin{minipage}{\textwidth}
\small
\textbf{Supplemntry Table 3.}
Machine learning predicted $T_C$ values for La-Zr-H compounds at various pressures.
The column with $\bar{e}$/H denotes the average number of electrons contributed per hydrogen atom. The column with HDOS fraction represents the hydrogen fraction of the total electronic density of states at the Fermi energy.
The column with range Mendeleev number denotes the range of the Mendeleev number of constituent elements. The column with mean Covalent indicates the covalent radii of constituent elements. The column with Avg. Dev. NValence represents the standard deviation of the number of valence electrons of constituent elements.
The first-principles calculated $T_C$ values using $\mu^*=0.10$ for the four selected systems ($R3m$-Zr$_2$H$_{17}$, $P\bar{6}m2$-LaZrH$_{18}$, $I4/mmm$-La$_2$ZrH$_{12}$, and $P6/mmm$-LaZr$_{2}$H$_{24}$) are shown in brackets. For the systems LaH$_{10}$, LaH$_{11}$, and LaH$_{23}$, the machine learning predicted $T_C$ values are listed first, followed by the first-principles DFT-calculated $T_C$ values in brackets, with references indicated as [1] and [2].\\[4pt]
\noindent\rule{\textwidth}{1.5pt}  
\resizebox{\textwidth}{!}{%
\begin{tabular}{l c c c c c c c c}
\toprule
\textbf{Formula} &
\textbf{Space group} &
\textbf{P (GPa)} &
\textbf{$\bar{e}$/H} &
\textbf{$T_C$ prediction (K)} &
\textbf{HDOS fraction} &
\textbf{Range Mendeleev Number} &
\textbf{Mean Covalent Radius} &
\textbf{Avg. Dev. NValence} \\
\hline
$\mathrm{La_3H_8}$ & Immm & 150 & 1.10 & 39.3 & 0.07 & 79.00 & 79.00 & 0.79 \\
$\mathrm{La_4H_{17}}$ & C2/m & 150 & 0.70 & 63.8 & 0.26 & 79.00 & 64.52 & 0.62 \\
$\mathrm{LaH_3}$ & Cmcm & 150 & 1 & 47.9 & 0.12 & 79.00 & 75.00 & 0.75 \\
$\mathrm{LaH_{11}}$ & P4/nmm & 150 & 0.27 & 159 (133~\cite{kruglov2020superconductivity}) & 0.35 & 79.00 & 45.67 & 0.31 \\
$\mathrm{LaH_{23}}$ & C2/m & 150 & 0.13 & 167 & 0.42 & 79.00 & 38.33 & 0.16 \\
$\mathrm{LaZrH_7}$ & P4/mmm & 150 & 1 & 56.7 & 0.14 & 79.00 & 66.56 & 0.86 \\
$\mathrm{Zr_3H_{14}}$ & Cmmm & 150 & 0.85 & 64 & 0.17 & 48.00 & 56.41 & 0.87 \\
$\mathrm{ZrH}$ & $R\bar{3}m$ & 150 & 4 & 16.9 & 0.02 & 48.00 & 103.00 & 1.50 \\
$\mathrm{ZrH_3}$ & Pm$\bar{3}$n & 150 & 1.33 & 53.8 & 0.10 & 48.00 & 67.00 & 1.13 \\
$\mathrm{ZrH_{14}}$ & C2/m & 150 & 0.28 & 81.7 & 0.35 & 48.00 & 40.60 & 0.37 \\
$\mathrm{La_3H_8}$ & Immm & 200 & 0.37 & 39.3 & 0.07 & 79.00 & 79.00 & 0.79 \\
$\mathrm{LaH_3}$ & Cmcm & 200 & 1 & 47.7 & 0.13 & 79.00 & 75.00 & 0.75 \\
$\mathrm{LaH_{10}}$ & Fm${3}$m & 200 & 0.3 & 261 (271~\cite{kruglov2020superconductivity}) & 0.41 & 79.00 & 47.00 & 0.33 \\
$\mathrm{LaH_{11}}$ & P4/nmm & 200 & 0.27 & 184 & 0.37 & 79.00 & 45.67 & 0.31 \\
$\mathrm{LaH_{23}}$ & C2/m & 200 & 0.13 & 170 (101~\cite{shutov2024ternary}) & 0.47 & 79.00 & 38.33 & 0.16 \\
$\mathrm{ZrH}$ & $R\bar{3}m$ & 200 & 4 & 17.0 & 0.03 & 48.00 & 103.00 & 1.50 \\
$\mathrm{ZrH_3}$ & $R\bar{3}c$ & 200 & 1.33 & 53.8 & 0.10 & 48.00 & 67.00 & 1.13 \\
$\mathrm{ZrH_4}$ & Fddd & 200 & 1 & 62.6 & 0.18 & 48.00 & 59.80 & 0.96 \\
$\mathrm{ZrH_6}$ & Cmc2$_1$ & 200 & 0.66 & 68.8 & 0.34 & 48.00 & 51.57 & 0.73 \\
$\mathrm{ZrH_{14}}$ & P4/nmm & 200 & 0.28 & 128.2 & 0.38 & 48.00 & 40.60 & 0.37 \\
$\mathrm{LaZr_{2}H_{24}}$ & P6/mmm & 200 & 0.46 & 229 (202) & 0.40 & 79.00 & 48.18 & 0.526 \\
$\mathrm{LaH_2}$ & Cmmm & 250 & 1.5 & 36.8 & 0.05 & 79.00 & 89.67 & 0.89 \\
$\mathrm{LaH_3}$ & Cmcm & 250 & 1 & 47.7 & 0.13 & 79.00 & 75.00 & 0.75 \\
$\mathrm{LaH_4}$ & I4/mmm & 250 & 0.75 & 60.7 & 0.24 & 79.00 & 66.20 & 0.64 \\
$\mathrm{LaH_{10}}$ & $R\bar{3}m$ & 250 & 0.3 & 261.5 & 0.42 & 79.00 & 47.00 & 0.33 \\
$\mathrm{LaH_{11}}$ & P4/nmm & 250 & 0.27 & 229.4 & 0.39 & 79.00 & 45.67 & 0.31 \\
$\mathrm{LaH_{23}}$ & C2/m & 250 & 0.13 & 170.0 & 0.50 & 79.00 & 38.33 & 0.16 \\
$\mathrm{ZrH_3}$ & $R\bar{3}c$ & 250 & 1.33 & 53.8 & 0.10 & 48.00 & 67.00 & 1.13 \\
$\mathrm{ZrH_6}$ & Cmc21 & 250 & 0.66 & 68.8 & 0.34 & 48.00 & 51.57 & 0.73 \\
$\mathrm{ZrH_{14}}$ & C2/m & 250 & 0.28 & 125.8 & 0.42 & 48.00 & 40.60 & 0.37 \\
$\mathrm{La_2ZrH_{12}}$ & I4/mmm & 250 & 0.83 & 71 (97) & 0.20 & 79.00 & 64.07 & 0.75 \\
$\mathrm{LaZr_2H_{12}}$ & I4/mmm & 250 & 0.91 & 51.7 & 0.13 & 79.00 & 61.93 & 0.85 \\
$\mathrm{LaH_2}$ & P6/mmm & 300 & 1.5 & 36.9 & 0.06 & 79.00 & 89.67 & 0.89 \\
$\mathrm{LaH_3}$ & Cmcm & 300 & 1 & 47.9 & 0.14 & 79.00 & 75.00 & 0.75 \\
$\mathrm{LaH_{11}}$ & P4/nmm & 300 & 0.27 & 230.6 & 0.40 & 79.00 & 45.67 & 0.31 \\
$\mathrm{LaH_{23}}$ & $P\bar{1}$ & 300 & 0.13 & 176.2 & 0.54 & 79.00 & 38.33 & 0.16 \\
$\mathrm{Zr_4H_{25}}$ & P1 & 300 & 0.64 & 83.5 & 0.30 & 48.00 & 50.86 & 0.71 \\
$\mathrm{ZrH_3}$ & $R\bar{3}c$ & 300 & 1.33 & 52.4 & 0.12 & 48.00 & 67.00 & 1.13 \\
$\mathrm{LaZrH_{18}}$ & P$\bar{6}$m2 & 300 & 0.38 & 244 (206) & 0.40 & 79.00 & 47.00 & 0.45 \\
$\mathrm{Zr_{2}H_{17}}$ & R3m & 300 & 0.47 & 187 (209) & 0.69 & 48.00 & 46.16 & 0.57 \\
\bottomrule
\end{tabular}
}
\end{minipage}
}
\end{table*}

\newpage
\clearpage

\bibliography{Reference-MS}
\bibliographystyle{naturemag}
\expandafter\ifx\csname url\endcsname\relax
  \def\url#1{\texttt{#1}}\fi
\expandafter\ifx\csname urlprefix\endcsname\relax\def\urlprefix{URL }\fi
\providecommand{\bibinfo}[2]{#2}
\providecommand{\eprint}[2][]{\url{#2}}
